\documentstyle[12pt]{article}
\renewcommand{\theequation}{\thesection.\arabic{equation}}

\def\tr{{\rm Tr}}
\def\a{\begin{eqnarray}}
\def\b{\end{eqnarray}}
\def\0{\nonumber}
\def\ba{\begin{array}}
\def\ea{\end{array}}
\def\noal{\noalign{\vskip10pt}}

\def\al{{\alpha}}
\def\lm{{\lambda}}

\def\tl{{\widetilde L}}

\def\hs{{\hat {\sigma}}}

\def\tm{{\widetilde M}}

\def\cm{{\cal M}}
\def\q{\bar {\cal Q}}
\newlength{\extraspace}
\newlength{\extraspaces}
\newcounter{dummy}
\newcommand{\ai}{
\addtocounter{equation}{1}
\setcounter{dummy}{\value{equation}}
\setcounter{equation}{0}
\renewcommand{\theequation}{\thesection.\arabic{dummy}\alph{equation}}
\begin{eqnarray}
\addtolength{\abovedisplayskip}{\extraspaces}
\addtolength{\belowdisplayskip}{\extraspaces}
\addtolength{\abovedisplayshortskip}{\extraspace}
\addtolength{\belowdisplayshortskip}{\extraspace}}
\newcommand{\bj}{
\end{eqnarray}
\setcounter{equation}{\value{dummy}}
\renewcommand{\theequation}{\thesection.\arabic{equation}}}
\def\d{{\partial}}

\newcommand{\ddlm}[1]{{\partial \over \partial \lm_{#1}}}

\newcommand{\ddg}[1]{{\partial \over \partial g_{#1}}}

\newcommand{\bac}{\begin{array}{c}}
\newcommand{\bacc}{\begin{array}{cc}}
\newcommand{\baccc}{\begin{array}{ccc}}
\newcommand{\barcl}{\begin{array}{rcl}}
\newcommand{\bacccc}{\begin{array}{cccc}}
\newcommand{\baccccc}{\begin{array}{ccccc}}
\newcommand{\baccccccc}{\begin{array}{ccccccc}}
\newcommand{\barclcrcl}{\begin{array}{rclcrcl}}
\newcommand{\bacl}{\begin{array}{cl}}
\newcommand{\bal}{\begin{array}{l}}
\newcommand{\bacll}{\begin{array}{cll}}
\def\noal{\noalign{\vskip10pt}}
\begin{document}
\begin{flushright}
SISSA-ISAS 84/94/EP\\
BONN-HE-08/94\\
hep-th/9407141
\end{flushright}
\vskip0.5cm
\centerline{\LARGE\bf Extended Toda lattice hierarchy,}
\vskip0.3cm
\centerline{\LARGE\bf extended two--matrix model}
\vskip0.3cm
\centerline{\LARGE \bf and $c=1$ string theory}
\vskip1cm
\centerline{\large  L.Bonora\footnote{ E-mail: bonora@tsmi19.sissa.it} }
\centerline{International School for Advanced Studies (SISSA/ISAS)}
\centerline{Via Beirut 2, 34014 Trieste, Italy}
\centerline{INFN, Sezione di Trieste.  }
\vskip0.5cm
\centerline{\large C.S.Xiong\footnote{E-mail: xiong@pib1.physik.uni-bonn.de}}
\centerline{Physikalisches Institut der Universit\"at Bonn}
\centerline{Nussallee 12, 53115 Bonn, Germany}
\vskip3cm
\abstract{We show how the two--matrix model and Toda lattice hierarchy
presented in a previous paper can be solved exactly: we obtain compact
formulas for correlators of pure tachyonic states
at every genus. We then extend the model to incorporate a set of discrete
states organized in finite dimensional $sl_2$ representations. We solve
also this extended model and find the correlators of the discrete states
by means of the $W$ constraints and the flow equations.
Our results
coincide with the ones existing in the literature in those cases in which
particular correlators have been explicitly calculated.
We conclude that the extented two--matrix model is a realization of the
discrete states of $c=1$ string theory.}

\vskip3cm
\vfill\eject

\section{Introduction}

Matrix models have been in the last few years the object of intense research.
This is particularly true for constant one--matrix models and
time--dependent one--matrix models. Considerably less attention has been paid
to two-- and multi--matrix models. We think instead two--matrix models
have a very rich structure and in fact provide a very convenient
description of what is known
as 2D gravity coupled to $c\leq 1$ matter.
In particular, in a recent paper \cite{BX1}
we showed that a two--matrix model underlies the
so--called $c=1$ string theory \cite{review}. We showed that the correlation
functions
of the pure {\it discrete tachyonic states} of the latter can be found in the
former and that the underlying integrable hierarchy is the same
for both. In \cite{BX1} we did not discuss however the full spectrum of
discrete states \cite{GKN},\cite{GK}\cite{P}. As is well--known, in the $c=1$
string theory there
appear {\it discrete states} labeled, say, by integers $r$ and $s$ (see below)
and
organized according to $sl_2$ finite dimensional representations;
the cases $r=0$ or $s=0$ correspond to the pure discrete tachyons
mentioned above. Occasionally we will call the discrete states with $r\neq0$
and $s\neq0$ simply {\it extra states}.

The present paper is a completion of \cite{BX1}. On the one hand
we show that the two--matrix model of that reference constitutes
a powerful tool to compute correlation functions in higher genera:
we exhibit in particular compact formulas for correlators at
arbitrary genus; these formulas are calculated in an amazingly
simple way.
On the other hand we show that the two--matrix model can
accommodate a full set of discrete states $\chi_{r,s}$ organized
according to the
finite dimensional representations of $sl_2$, exactly like the
discrete states of the $c=1$ string theory. To this end we have to modify the
two--matrix model discussed in \cite{BX1} by adding new interaction terms
(like in \cite{BX2}).
The $W$--constraints of the modified model turn out to be a very efficient
tool to calculate the correlation functions of the discrete states at any
genus.
Whenever corresponding results in $c=1$ string theory are available in the
literature we show that
they coincide with ours -- but our results are much more
general.  As a consequence we
consider our modified two--matrix model as a realization
of the discrete states of $c=1$ string theory.

The paper is organized as follows.
In section 2 we present once again the model of \cite{BX1} and define its
extension. We then present the corresponding integrable hierarchy (the
extended Toda lattice hierarchy) and the corresponding $W$ constraints.
In section 3 we define our discrete states and their $sl_2$ representation
properties. We then start the program of calculating their correlation
functions
(CF's) in a definite small phase space, defined at the beginning of section
3.2:
after some general theorems we concentrate on genus 0 correlators
and give a few compact formulas. Section 4 is devoted to a discussion of
a dispersionless Toda hierarchy, which underlies the previous calculations.
In section 5 we show the connection with the
Penner model. In section 6
we return to CF's in higher genera and derive the compact formulas announced
above. In section 7 we discuss the model in a larger phase space.
Appendix A is devoted to the discussion of the subtleties of the continuum
limit, while in Appendix B we collect a few elements in favor of a topological
field theory interpretation of $c=1$ string theory.

\section{The extended two--matrix model}

\setcounter{equation}{0}
\setcounter{subsection}{0}

The model of two Hermitean $N \times N$ matrices $M_1$ and $M_2$,
considered in \cite{BX2},\cite{BX1}, was introduced in terms of the partition
function
\a
Z_N(t,c)=\int dM_1dM_2 e^{TrU},\quad\quad
U=V_1 + V_2 + g M_1 M_2\label{Zo}
\b
with potentials
\a
V_{\al}=\sum_{r=1}^{\infty}t_{\al,r}M_{\al}^r\,\qquad \al=1,2.\label{V}
\b
Although nothing prevent us from solving this model with $N$ finite,
the spirit of this paper is to eventually take $N\to \infty$; the renormalized
$N$ will be called $x$ and identified with the cosmological constant.

After integration of the angular variables eq.(\ref{Zo}) can be written as
\a
Z_N(t,C)&=&{\rm const} \int \prod_{\alpha=1}^2 \prod_{i=1}^N
d\lambda_{\alpha,i}
\Delta (\lambda_1)\Delta(\lambda_2)
\exp  U,\label{Z}\\
U&=&\sum_{k=1}^{\infty} t_{1,k}\sum_{i=1}^N
\lambda_{1,i}^k+
\sum_{r=1}^{\infty} t_{2,r}\sum_{i=1}^N \lambda_{2,i}^r
+g\sum_{i=1}^N \lambda_{1,i}\lambda_{2,i}\label{U}
\b
where $\Delta$ stands for the Vandermonde determinant.
We refer to this model as the {\it ordinary two--matrix model}.
This model can only accommodate the pure tachyonic states. If we
want to describe the other discrete states we have to enlarge
the model in the same way as we did in section 4
of \cite{BX2}, that is we add new interaction terms as follows
\a
&&U \rightarrow {\cal U}=\sum_{k=1}^{\infty} t_{1,k}\sum_{i=1}^N
\lambda_{1,i}^k +\sum_{r=1}^{\infty} t_{2,r}\sum_{i=1}^N \lambda_{2,i}^r
+{\cal V},\quad\quad
{\cal V} = \sum_{k,r\geq 1}g_{k,r} \sum_{i=1}^N
 \lambda_{1,i}^k \lambda_{2,i}^r\0
\b
where $g_{1,1} \equiv g$.
We notice that this corresponds, in general, to adding to
the original model (\ref{Zo}) not terms $Tr(M_1^k M_2^r)$,
but rather composite operators of the form $Tr(D_1^kD_2^r)$,
where $D_1$ and $D_2$ are the diagonal eigenvalue matrices
of $M_1$ and $M_2$, respectively.

We will refer to this new model as to the {\it extended two--matrix model.}
We can generalize to this model everything we did for the original
one (\ref{Zo}). In particular we can map this functional integral problem
into a linear integrable system together with definite coupling constraints.
In the following we will not repeat the procedure (see \cite{BX2}),
but simply write down definitions and results. First let us remind
our conventions. For any matrix $M$, we define
\a
\bigl(\cm\bigl)_{ij}= M_{ij}{{h_j}\over{h_i}},\qquad
{\bar M}_{ij}=M_{ji},\qquad
M_l(j)\equiv M_{j,j-l}.\0
\b
As usual we introduce the natural gradation
\a
deg[E_{ij}] = j -i\0
\b
and, for any given matrix $M$, if all its non--zero elements
have degrees in the interval $[a,b]$, then we will simply
write: $M\in [a,b]$. Moreover $M_+$ will denote the upper triangular
part of $M$ (including the main diagonal), while $M_-=M-M_+$. We will write
${\rm Tr} (M)= \sum_{i=0}^{N-1} M_{ii}$. Finally, for the sake of
compactness, we will often use the convention
\a
t_{1,r} \equiv g_{r,0}, \qquad\qquad t_{2,s} \equiv g_{0,s}\0
\b

As usual we introduce the orthogonal polynomials
\a
\xi_n(\lambda_1)=\lambda_1^n+\hbox{lower powers},\qquad\qquad
\eta_n(\lambda_2)=\lambda_2^n+\hbox{lower powers}\0
\b
and the orthogonal functions
\a
\Psi_n(\lambda_1)=e^{V_1(\lambda_1)}\xi_n(\lambda_1),
\qquad
\Phi_n(\lambda_2)=e^{V_2(\lambda_2)}\eta_n(\lambda_2).\0
\b
The orthogonality relations are
\a
\int d\lm_1 d\lm_2\Psi_n(\lambda_1)e^{{\cal V}(\lambda_1 ,\lambda_2)}
\Phi_m(\lambda_2)=\delta_{nm}h_n(t,g).\label{orth}
\b
In these equations
\a
V_\al(\lambda_\al) = \sum_{r=1}^\infty t_{\al,r}\lambda_{\al}^r, \qquad \al=
1,2,
,\qquad  {\cal V}(\lambda_1,\lambda_2) =
\sum_{k,r\geq 1}g_{k,r} \lambda_{1}^k \lambda_{2}^r\0
\b
Using these relations and the properties
of the Vandermonde determinants, one can easily
calculate the partition function
\a
Z_N(t,g)={\rm const}~N!\prod_{i=0}^{N-1}h_i\label{parti1}
\b
Knowing the partition function means knowing
the coefficients $h_n(t,g)$.

We will denote the semi--infinite column vectors with components
$\Psi_0,\Psi_1,\Psi_2,\ldots,$ and  $\Phi_0,\Phi_1,$ $\Phi_2,\ldots,$
by $\Psi$ and $\Phi$, respectively.

Next we introduce the following $Q$-- and $P$--type matrices
\ai
&&\int d\lm_1 d\lm_2\Psi_n(\lambda_1)
\lm_{\al}e^{{\cal V}(\lm_1,\lm_2)}
\Phi_m(\lambda_2)\equiv Q_{nm}(\al)h_m,\quad
\al=1,2.\label{Q}\\
&&\int d\lm_1 d\lm_2\Bigl(\ddlm 1 \Psi_n(\lambda_1)\Bigl)
e^{{\cal V}(\lm_1,\lm_2)}\Phi_m(\lambda_2)\equiv P_{nm}(1)h_m\label{P(1)}\\
&&\int  d\lambda_1d\lambda_2\Psi_n(\lambda_1)e^{{\cal V}(\lm_1,\lm_2)}
\Bigl(\ddlm 2 \Phi_m(\lambda_2)\Bigl)\equiv P_{mn}(2)h_n\label{P(2)}
\bj
Both $Q(1)$ and $\bar {\cal Q}(2)$ are
Jacobi matrices: their pure upper triangular
part is $I_+=\sum_i E_{i,i+1}$.

Here come now the three basic elements of our analysis. The {\it first}
is provided by the following coupling constraints
\a
P(1)+\sum_{r,s\geq 1} rg_{r,s}Q(1)^{r-1}Q(2)^s =0,\quad\quad
\bar{\cal P}(2) +\sum_{r,s\geq 1} sg_{r,s}Q(1)^{r}Q(2)^{s-1}=0
\label{coupling}
\b
These coupling conditions lead to the $W_{1+\infty}$--constraints (see
below). {}From them it follows at once that
\a
Q(1)\in[-\infty, 1],\qquad Q(2) \in [-1 , \infty]\0
\b

The {\it second element} are the discrete linear systems
associated to the extended two--matrix model.
The {\it first discrete linear system} is
\a
\left\{\ba{ll}
Q(1)\Psi(\lambda_1)=\lambda_1\Psi(\lambda_1),& \\\noal
{\partial\over{\partial g_{r,0}}}\Psi(\lambda_1)=Q^r_+(1)
\Psi(\lambda_1),\qquad\qquad r\geq 1&\\\noal
{\partial\over{\partial g_{r,s}}}\Psi(\lambda_1)=-\Big(Q^r(1)Q^s(2)\Big)_-
\Psi(\lambda_1),\quad\quad r\geq 0,~~s\geq 1&\\\noal
{\partial\over{\partial\lm}}\Psi(\lambda_1)=P(1)\Psi(\lm_1).&
\ea\right.\label{DLS1}
\b
The corresponding consistency conditions are
\ai
&&[Q(1), ~~P(1)]=1, \label{CC11}\\
&&{\partial\over{\partial g_{r,s}}}Q(1)=[Q(1), \Big(Q^r(1)Q^s(2)\Big)_-],
\qquad r\geq 0,~s\geq 1,~r+s\geq 1\label{CC12}\\
&&{\partial\over{\partial g_{r,0}}}P(1)=[Q^r_+(1), P(1)],\qquad
\qquad r\geq 1\label{CC13}\\
&&{\partial\over{\partial g_{r,s}}}P(1)=[P(1), \Big(Q^r(1)Q^s(2)\Big)_-],
\quad\quad r\geq 0,~s\geq 1\label{CC14}
\bj

The {\it second discrete linear system} is
\a
\left\{\ba{ll}
\bar {\cal Q}(2)\Phi(\lambda_2)=\lambda_2\Phi(\lambda_2),& \\\noal
{\partial\over{\partial g_{0,s}}}\Phi(\lambda_2)=\bar {\cal Q}^s_+(2)
\Phi(\lambda_2),\qquad \qquad s\geq 1&\\\noal
{\partial\over{\partial
g_{r,s}}}\Phi(\lambda_2)=-\Big(\bar {\cal Q}^s(2)\bar{\cal Q}^r(1)\Big)_-
   \Phi(\lambda_2), \qquad r\geq 1,s\geq 0&\\\noal
{\partial\over{\partial\lm}}\Phi(\lambda_2)=P(2)\Psi(\lm_2).&
\ea\right.\label{DLS2}
\b
The corresponding consistency conditions are
\ai
&&[\bar {\cal Q}(2), ~~P(2)]=1\label{CC21}\\
&&{\partial\over{\partial g_{r,s}}} Q(2)=
[ \Big(Q^r(1)Q^s(2)\Big)_+,~~ Q(2)],\qquad r\geq 1,~s\geq 0, ~r+s \geq 1
 \label{CC22}\\
&&{\partial\over{\partial g_{r,s}}}P(2)=
[P(2),\Big(\bar {\cal Q}^s(2)\bar{\cal Q}^r(1)\Big)_-], \qquad r\geq 1,
s\geq 0\label{CC23}\\
&&{\partial\over{\partial g_{0,s}}}P(2)=[\bar {\cal Q}^s_+(2), P(2)],\qquad
s\geq 1 \label{CC24}
\bj

The {\it third ingredient} we need is the link between the
quantities that appear in the linear systems and in the coupling conditions
with the original partition function. We have
\a
{\d \over {\d g_{r,s}}} \ln Z_N(t,g) = {\rm Tr} \Big(Q^r(1)Q^s(2)\Big),
\label{ddZ}
\b
It is evident that, by using the above flow equations, we can express all
the derivatives of $Z_N$ in terms of the elements of the $Q$ matrices.
For example
\a
{\d^2\over{\d t_{1,1}\d t_{\al,r}}}
\ln Z_N(t,g)=\Bigl(Q^r(\al)\Bigl)_{N,N-1},\quad \al =1,2\label{parti3}
\b
Knowing all the derivatives with respect to the coupling parameters
we can reconstruct the partition function up to an overall integration
constant (depending only on $N$).

The three elements just introduced provide a larger definition of
our system. In fact we
notice that the initial path integral exists only in a very restricted
region of the parameter space. The above formulas, whenever they are
meaningful,
provide the continuation of our system to regions of the parameter space
where the path--integral is ill--defined.

We will use the following coordinatization for $Q(1)$ and $Q(2)$
\a
Q(1)=I_++\sum_i \sum_{l=0}^\infty a_l(i)E_{i,i-l}, \qquad\qquad\qquad
\q(2)=I_++\sum_i \sum_{l=0}^\infty b_l(i)E_{i,i-l}\label{jacobi}
\b
One can immediately see that, for example,
\a
\Bigl(Q_+(1)\Bigl)_{ij}=\delta_{j,i+1}+a_0(i)\delta_{i,j},\qquad
\Bigl(Q_-(2)\Bigl)_{ij}=R_i\delta_{j,i-1}\label{jacobi1}
\b
where $R_{i+1} \equiv h_{i+1}/h_i$.
As a consequence of this coordinatization, eq.(\ref{parti3}) gives in
particular the important relation
\a
 {\d^2\over{\d t_{1,1}\d t_{2,1}}}\ln Z_N(t, g) = R_N \label{ZR}
\b
and
\a
{{\d^2 }\over{\d t_{1,1} \d t_{2,1}}} \ln R_j = R_{j+1}-2R_j
+R_{j-1}\label{Todaeq}
\b

In our model, the four matrices $Q(1), Q(2)$ and $P(1),P(2)$ are not totally
independent. In fact the two $P$ matrices can be expressed in terms of
the two $Q$ Jacobi matrices as follows
\a
&& P_{nm}(1) = \sum_{r=1}^\infty rt_{1,r} Q^{r-1}_{nm}(1) + n Q^{-1}_{nm}(1) +
\sum_{r=1}^\infty {{\d \ln Z_n}\over{\d t_{1,r}}} Q^{-r-1}_{nm}(1);
\label{P1}\\
&& {\bar{\cal P}}_{nm}(2)
= \sum_{r=1}^\infty rt_{2,r} Q^{r-1}_{nm}(2) + m Q^{-1}_{nm}(2) +
 \sum_{r=1}^\infty Q^{-r-1}_{nm}(2) {{\d \ln Z_m}\over{\d t_{2,r}}}. \label{P2}
\b
It is worth proving these formulas in some detail. Let us see the first
equation
as an example. From the spectral equation in the first discrete linear
system, we see that the polynomials $\xi_n(\lambda_1)$ satisfy the recursion
relation
\a
\lambda_1\xi_n(\lambda_1) = \sum_{m=0}^{n+1}Q(1)_{nm}\xi_m(\lambda_1) \0
\b
Solving this recursion relation we get
\a
\xi_n(\lambda_1) ={\rm det} \Bigl(\lambda_1-Q(1)\Bigl)_{n\times n} \0
\b
The determinant on the RHS refers to the principal minor delimited by
(and including) the n--th row and n--th column. Equivalently
\a
\ln\xi_n(\lambda_1) = n\ln \lambda_1 -\sum_{r=1}^\infty \frac{1}{l\lambda_1^r}
\frac{\d\ln Z_n}{\d t_{1,r}} \0
\b
This implies
\a
\ln\Psi_n(\lambda_1) = \sum_{r=1}^\infty t_{1,r} \lambda_1^r +
n\ln \lambda_1 - \sum_{r=1}^\infty \frac{1}{r\lambda_1^r}
\frac{\d\ln Z_n}{\d t_{1,r}} \label{psin}
\b
Taking the derivative with respect to $\lambda_1$, and using
the spectral relation, we have
\a
\frac{\d \Psi_n(\lambda_1)}{\d \lambda_1}
 &=&\Bigl( \sum_{r=1}^\infty rt_{1,r} \lambda_1^{r-1} + n \lambda_1^{-1} +
 \sum_{r=1}^\infty {{\d \ln Z_n}\over{\d t_{1,r}}} \lambda_1^{-r-1} \Bigl)
 \Psi_n(\lambda_1) \0\\
 &=&\Bigl( \sum_{r=1}^\infty rt_{1,r} Q^{r-1}_{nm}(1) + n Q^{-1}_{nm}(1) +
 \sum_{r=1}^\infty {{\d \ln Z_n}\over{\d t_{1,r}}} Q^{-r-1}_{nm}(1) \Bigl)
 \Psi_m(\lambda_1) \0
\b
{}From the last equality we read off formula (\ref{P1}). Similarly
we obtain
\a
\ln\Phi_n(\lambda_2) = \sum_{r=1}^\infty t_{2,r} \lambda_2^r +
n\ln \lambda_2 - \sum_{r=1}^\infty \frac{1}{r\lambda_2^r}
\frac{\d\ln Z_n}{\d t_{2,r}} \label{phin}
\b
from which we can extract (\ref{P2}).

Among the two sets of flow equations
eqs.(\ref{CC12}-\ref{CC14}) and eqs.(\ref{CC22}-\ref{CC24}), we
can choose as independent the following two
\ai
&&{\partial\over{\partial g_{r,s}}}Q(1)=[Q(1),~~\Bigl(Q^r(1)Q^s(2)\Bigl)_-],
\label{extoda1}\\
&&{\partial\over{\partial g_{r,s}}}Q(2)=[\Bigl(Q^r(1)Q^s(2)\Bigl)_+, ~~Q(2)].
 \label{extoda2}
\bj
It is easy to see that if we switch off the new extra couplings (i.e.
we set  $g_{r,s}$ to zero except $g_{0,s}$, $g_{r,0}$ and $g_{1,1}$), we
recover
the usual 2--dimensional Toda lattice hierarchy. Therefore the above
equations are an extension of Toda lattice hierarchy. Their integrability
is guaranteed by the commutativity of the flows.

\subsection{$W_{1+\infty}$ constraints}

The $W_{1+\infty}$ constraints (or simply $W$--constraints)
on the partition function for the extended two--matrix
model were obtained in \cite{BX2}.
They take the form
\ai
&&\Big({\cal L}^{[r]}_n(1)-(-1)^rT^{[r]}_n(1)\Big)Z_N(t;g)=0,\qquad
r\geq1;\qquad n\geq-r.
\label{W1}\\
&&\Big({\cal L}^{[r]}_n(2)-(-1)^r T^{[r]}_n(2)\Big)Z_N(t;g)=0,
\qquad r\geq1, \qquad n\geq -r\label{W2}
\bj
The generators ${\cal L}^{[r]}_n(1)$ are differential operators involving
$N$ and $t_{1,k}$, while ${\cal L}^{[r]}_n(2)$ have the same form
with $t_{1,k}$ replaced by $t_{2,k}$. The $T^{[r]}_n(1)$ are differential
operators involving the couplings $g_{r,s}$ (except $g_{r,0}$) and
commute with the ${\cal L}^{[r]}_n(1)$. The $T^{[r]}_n(2)$ are differential
operators involving the couplings $g_{r,s}$ (except $g_{0,s}$) and
commute with the ${\cal L}^{[r]}_n(2)$.

The ${\cal L}^{[r]}_n(1)$ satisfy a $W_{1+\infty}$ algebra
\a
[{\cal L}^{[r]}_n(1), {\cal L}^{[s]}_m(1)]=(sn-rm)
 {\cal L}^{[r+s-1]}_{n+m}(1)+\ldots,\label{LLgen}
\b
for $r,s\geq 1;~n\geq-r,m\geq-s$. Here dots denote lower than $r+s-1$ rank
operators.
The algebra of the ${\cal L}^{[r]}_n(2)$ is just a copy of the above one,
and is isomorphic to the algebra satisfied by the $T^{[r]}_n(1)$ and by the
$T^{[r]}_n(2)$ separately.

There is a compact way to write down the above generators. It consists
in introducing the current
\a
J(z) = \sum_{r=1}^\infty r t_r z^{r-1} + {N \over z} + \sum_{r=1}^\infty
z^{-r-1} {\d \over {\d t_r}}\label{J}
\b
and defining the density
\a
{\cal L}^{[n]}(z) = {1\over{n+1}} :\Bigl( -\d + J(z)\Bigl)^{n+1}: \cdot 1
\label{J1}
\b
Then ${\cal L}_k^{[n]}$ can be recovered as
\a
{\cal L}_k^{[n]} = {\rm Res}_{z=0} ({\cal L}^{[n]}(z) z^{n+k})\label{J2}
\b

The above definition holds for both the 1 and 2 sector. As for the
$T^{[r]}_n$ operators we proceed as follows. We define the bilocal densities
\a
&&H_1(z,w)= \sum_{r,s\geq 0} rg_{r,s} z^{r-1}w^{s-1},\qquad
H_2(z,w)= \sum_{r,s\geq 0} sg_{r,s} z^{r-1}w^{s-1},\0\\
&&K(z,w) = \sum_{r,s\geq 0} z^{-r-1}w^{-s-1} {\d\over{\d g_{r,s}}}\0
\b
where $r$ and $s$ are never simultaneously 0, and
\a
{\cal W}^{[r]}_1= : K \Big( w^{-1} \d_z + H_1\Big)^r:\cdot 1,\qquad
{\cal W}^{[r]}_2= : K \Big( z^{-1} \d_w + H_2\Big)^r:\cdot 1\label{W12}
\b
Then
\ai
&&T^{[r]}_n(1) = {\rm Res}_{z=0} {\rm Res}_{w=0}
\Big({\cal W}^{[r]}_1z^{r+n}w^r\Big)
\label{T1}\\
&&T^{[r]}_n(2) = {\rm Res}_{z=0} {\rm Res}_{w=0}
\Big({\cal W}^{[r]}_2z^{r}w^{r+n}\Big)
\label{T2}
\bj
For example
\a
T^{[1]}_n(1)=\sum_{r,s\geq 1}rg_{r,s}
{\partial\over{\partial g_{r+n,s}}},\qquad
T^{[1]}_n(2)=\sum_{r,s\geq 1}sg_{r,s}
{\partial\over{\partial g_{r,s+n}}},\qquad n\geq -1\0
\b

\section{Discrete states and their correlation functions}

\setcounter{equation}{0}
\setcounter{subsection}{0}

The purpose of this section is to show that the extended two--matrix model
can accomodate {\it discrete states} organized in finite dimensional $sl_2$
representations, analogous to the $c=1$ string theory ones,
and to calculate their correlation
functions via the $W$--constraints introduced before.

\subsection{Definition and properties of the discrete states}

{\it We call discrete states the operators $\chi_{r,s}$ coupled
to the $g_{r,s}$}.  In this definition $r=s=0$ is excluded. However, for
later use we introduce also $\chi_{0,0}\equiv Q$ as the operator coupled to
$g_{0,0}\equiv N$.
{}From the very definition it is apparent that, classically, $\chi_{r,s}$ is
represented by $\sum_{i=1}^N\lambda_i^r \mu_i^s$, where, to simplify our
notation we set $\lambda_1=\lambda$ and $\lambda_2=\mu$. As a
consequence these states carry a built--in  $sl_2$ structure.
For let us define
\a
H={1\over 2}\sum_{i=0}^N\Big(\lm_i {\d \over {\d \lm_i}}- \mu_i {\d \over
{\d \mu_i}}\Big),\qquad E_+ =\sum_{i=1}^N \lm_i {\d \over {\d
\mu_i}},\qquad E_- = \sum_{i=1}^N \mu_i {\d \over {\d \lm_i}}\0
\b
Then
\a
\relax[H,E_\pm] = \pm E_\pm, \qquad [E_+ , E_-]= 2H \0
\b
and
\a
H \chi_{r,s} = {1\over 2}(r-s)\chi_{r,s},\quad\quad E_+ \chi_{r,s} =
s \chi_{r+1,s-1},\quad\quad E_- \chi_{r,s} = r \chi_{r-1,s+1}\0
\b
Therefore the set $\{\chi_{r,s}=\sum_{i=1}^N \lm_i^r \mu_i^s,~~ r+s=n\}$ form
an
(unnormalized) representation of this algebra of dimension $n+1$.

We can do even better and introduce the new states
\a
\omega_{r,s} = \sqrt {{(r+s)!}\over {r!s!}}\chi_{r,s}\0
\b
endowed with the scalar product
\a
(\omega_{r,s},\omega_{r',s'}) = \delta_{r,r'}\delta_{s,s'}\0
\b
One can easily verify that
\a
\omega_{r,s}=|j,m>, \quad\quad r=j+m,\quad\quad s=j-m\0
\b
is the standard basis for finite dimensional $sl_2$ representations.
It is also easy to see that the products $\chi_{r,s} \chi_{r',s'}$
with $r+s=2j$, $r'+s'=2j'$ span the tensor product of the representations
$j$ and $j'$. It can be therefore decomposed into irreducible
representations
\a
\omega_{r_1,s_1}\omega_{r_2,s_2} = \sum_{r,s} C(r_1,s_1,r_2,s_2|r,s)
\omega_{r,s}\label{tensorpr}
\b
where the summation extends to the range $M(r_1-s_2,r_2-s_1)\leq r \leq
r_1+r_2$
and $M(s_1-r_2, s_2-r_1)\leq s \leq s_1+s_2$. $M(x,y)$ means the maximum
between $x$ and $y$.
Here $C$ are the standard Clebsh--Gordan coefficients expressed in terms of
the labels $r,s$ instead of $j,m$. A specification is in order:
when $j=j'$ the symmetric representations of the Clebsh--Gordan series
are absent in these classical products.

To end this introduction let us recall that $\chi_{r,0}$ and $\chi_{0,s}$
coincide with the operators $\tau_r$ and $\sigma_s$
introduced in \cite{BX1},respectively, which, in turn, were identified with
the purely tachyonic states ${\cal T}_r$ and ${\cal T}_{-s}$ of the $c=1$
string theory. It is therefore natural to try to interpret the $\chi_{r,s}$
as representatives of the discrete states of the $c=1$ string theory.

\subsection{Correlation functions of the discrete states. Generalities}

The correlation functions of the extended two--matrix model are
in general defined by
\a
\ll \chi_{r_1,s_1}\ldots \chi_{r_k,s_k}\gg =
{\partial\over{\partial g_{r_1,s_1}}}
 \ldots {\partial\over{\partial g_{r_k,s_k}}} \ln Z_N\0
\b

Our main purpose in this paper is to calculate the correlation functions in
a very simple {\it small phase space} ${\cal S}_0$: we set $g_{i,j}=0$
except for $g_{1,1}\equiv g$, which is left undetermined (but in the
topological
application we will set $g=-1$).
As a consequence the CF's
will be functions of $g$ and $N$. We denote by the symbol $<\cdot>$ the
CF's calculated is such a small phase space \footnote{In the following we often
use the expressions {\it the model ${\cal S}_0$ or the critical point
${\cal S}_0$} as equivalent to {\it the small phase space ${\cal S}_0$}.}.

In order to calculate the correlation functions we have simply
to write down the $W$ constraints in terms of them. We obtain a set of
(overdetermined)
algebraic equations which in general one can solve recursively. Let us
see first of all two very general lemmas.

{\bf Lemma 1}. The CF satisfy the symmetry property
\a
<\chi_{r_1,s_1}\ldots\chi_{r_n,s_n}>= <\chi_{s_1,r_1}\ldots\chi_{s_n,r_n}>
\label{sym}
\b
This is due to the symmetry of the $W$ constraints (\ref{W1}) and
(\ref{W2}) and of the small phase space under the exchange
$1\leftrightarrow 2$.

{\bf Lemma 2}. In the above defined small phase space the CF's satisfy
\a
(r_1+\ldots +r_n- s_1-\ldots -s_n)<\chi_{r_1,s_1}\ldots\chi_{r_n,s_n}>=0
\label{lemma}
\b

In order to prove this lemma one rewrites the $W$ constraints (\ref{W1})
and (\ref{W2}) with $r=1$ and $n=0$ as follows
\a
\sum_{r\geq 1, s\geq 0} rg_{r,s} <\chi_{r,s}> + {1\over 2}
N(N+1) =0\0\\
\sum_{r\geq 0, s\geq 1} sg_{r,s} <\chi_{r,s}> + {1\over 2}
N(N+1) =0\0
\b
respectively. One differentiates these equations w.r.t.
$g_{r_1,s_1}$, \ldots,
$g_{r_n,s_n}$ and sets $g_{r,s}=0$ except $g_{1,1}= g$. One gets
\a
(r_1 +\ldots +r_n)<\chi_{r_1,s_1}\ldots \chi_{r_n,s_n}> + g
<\chi_{1,1}\chi_{r_1,s_1}\ldots \chi_{r_n,s_n}>=0\0\\
(s_1 + \ldots +s_n) <\chi_{r_1,s_1}\ldots \chi_{r_n,s_n}> +g
<\chi_{1,1}\chi_{r_1,s_1}\ldots \chi_{r_n,s_n}> =0\0
\b
Subtracting the second equation from the first we obtain the result.

This lemma reflects the $sl_2$ structure of the discrete states since it
means nothing but the conservation of the $H$ eigenvalue (conservation of
the third angular momentum component).

Consider the $W$ constraints
\a
\Big({\cal L}^{[r]}_0(1)-(-1)^rT^{[r]}_0(1)\Big)Z_N(t;g)=0,\qquad r\geq1
\label{Wr0}
\b
and evaluate it in ${\cal S}_0$. We get
\a
&&{\cal L}^{[r]}_0(1) Z_N = {1\over {r+1}} N(N+1)\ldots(N+r)Z_N\label{LZ}\\
&&T^{[r]}_0(1)Z_N= g^r <\chi_{r,r}>Z_N\label{TZ}
\b
This allows us to calculate $<\chi_{r,r}>$ for any genus. Remember that
the genus $h$ and genus $h-1$ contributions differ by $N^2$. Therefore
the genus $h$ contribution, $<\chi_{r,r}>_h$, is given by
\a
<\chi_{r,r}>_h= \left\{\bacc{N^{r+1-2h} \over {(r+1) (-g)^r}} b_{2h}(r)
\qquad &r> 2h-1\\
0 \qquad& r\leq 2h-1\ea\right.\label{chinn}
\b
where
\a
b_k(n) =\sum_{1\leq r_1<r_2<\ldots<r_k\leq n} r_1r_2 \ldots
r_k,\qquad\qquad b_0(n) =1
 \label{akn}
\b

Other exact results will be obtained in section 5. From now on in
this section we concentrate on calculating genus zero CF's.

\subsection{Correlation functions. Genus 0 results.}

In genus 0 (alias dispersionless limit) one can produce very general compact
formulas for the CF's. In order to find the $W$ constraints appropriate for
genus 0 we proceed as follows (see Appendix). We assign a
homogeneity degree $[\cdot]$ to each of the involved quantities
\a
\relax[g_{r,s}] = 1, \qquad [N]=1,\qquad [\ln Z^{(0)}] = 2 \0
\b
where the superscript $^{(0)}$ denotes
the genus 0 contribution. Then we keep only the leading terms in this
degree in the $W$ constraints. Henceforth we denote $N$ by $x$ (for a
motivation
see Appendix A and end of section 4); the latter will be associated later
on to the cosmological constant. We will
denote the genus zero part of every CF $<\cdot>$ by $<\cdot>_0$.

The simplest result and the first ingredient we need is $<\chi_{n,n}>_0$.
This CF we have calculated exactly above, (\ref{chinn}). The genus 0
contribution is
\a
<\chi_{n,n}>_0= {x^{n+1} \over {(-g)^n}} {1\over
{n+1}}\label{chinn0}
\b
For the remaining CF's the relevant $W$
constraints are
\ai
&&{\cal L}^{[r]}_{-r}(1)Z(t;g,x)=(-1)^rT^{[r]}_{-r}(1)Z(t;g,x),\qquad
r\geq1 \label{W1r-r}\\
&&{\cal L}^{[r]}_n(1)Z(t;g,x)=(-1)^r
T^{[r]}_n(1)Z(t;g,x), \qquad r\geq1, \qquad n\geq 1\label{W1rn}
\bj
and the analogous ones with $1\rightarrow 2$. But we will not have
to consider the latter due to Lemma 1. Performing the degree analysis one
easily extracts the dispersionless limit. In particular we have
\a
T_n^{[r]}(1)= \sum_{\stackrel {i_1,\ldots,i_r\geq 1}
{j_1,\ldots,j_r\geq 1}}i_1\ldots i_r g_{i_1,j_1}\ldots
g_{i_r,j_r} \frac{\d}{\d g_{i_1+\ldots+i_r+n,j_1+\dots+j_r} }.\label{tnr0}
\b
These generators satisfy the algebra
\a
[T^{[r]}_n(1), T^{[s]}_m(1)]=(sn-rm)T^{[r+s-1]}_{n+m}(1),
\qquad r,s\geq 1;\quad n\geq-r,\quad m\geq-s \label{areapres}
\b
which characterizes the area--preserving diffeomorphisms.
The algebra of the $T^{[r]}_n(2)$ is just a copy of the above one.
Similar simplified (but not quite as simple)
expressions can be gotten for the ${\cal L}$--type generators.

Now, from (\ref{W1r-r}) and (\ref{W1rn}), respectively, we obtain
\a {\rm LHS}=\sum_{\stackrel
{k_1,\ldots,k_r\geq 1}{l_1,\ldots, l_r\geq 1}}k_1\ldots k_r g_{k_1,l_1}\ldots
g_{k_r,l_r}<\chi_{k_1+\ldots+k_r-r, l_1+\ldots+l_r}>_0
\label{W1r-r0}
\b
and
\a
&&\sum_{l=1}^{r+1} {1\over{r+1}} \left(
\begin{array}{c} r+1\\ l\end{array}\right) x^{r-l+1} \sum_{k_1+\ldots +k_l
= n} <\chi_{k_1,0}>_0\ldots<\chi_{k_l,0}>_0=\label{W1rn0}\\
&&~~~~~~~~~~~~~~~~(-1)^r \sum_{\stackrel {k_1,\ldots,k_r\geq 1}
{l_1,\ldots,l_r\geq 1}}k_1\ldots k_r g_{k_1,l_1}\ldots
g_{k_r,l_r}<\chi_{k_1+\ldots+k_r+n, l_1+\ldots+l_r}>_0\0
\b
 The LHS in
(\ref{W1r-r0}) needs not be written down explicitly (see below).

The first step consists in differentiating (\ref{W1r-r0}) with respect to
$g_{n,n-r}$ with $n>r$ and evaluating the result in the small phase space
${\cal S}_0$. One gets
\a
g^r <\chi_{0,r} \chi_{n,n-r}>_0 + r n g^{r-1} <\chi_{n-r,n-r}>_0 =0\0
\b
since the LHS of (\ref{W1r-r0}) does not depend on $g_{n,n-r}$ with $n>r$.
Inserting (\ref{chinn0}) we obtain
\a
<\chi_{0,r} \chi_{n,n-r}>_0= {x^n\over {(-g)^n}}r=
<\chi_{r,0}\chi_{n-r,n}>_0\label{chi0r}
\b
The second equality follows from Lemma 1. Now differentiate (\ref{W1rn0})
w.r.t. $g_{n,n+p}$ with $p>0$, and evaluate the result in ${\cal S}_0$.
One gets
\a
x^r <\chi_{p,0}\chi_{n,n+p}>_0 &=& (-1)^r \Big(rn
g^{r-1}<\chi_{n+p+r-1,n+p+r-1}>_0\0\\
&&~~~~~~~~+g^r <\chi_{p+r,r}\chi_{n,n+p}>_0\Big)\label{chinn+p}
\b
for only the first terms in the LHS of (\ref{W1rn0}) contributes, due to
Lemma 2. Finally, using (\ref{chi0r}) and (\ref{chinn0}) one gets
\a
<\chi_{p+r, r}\chi_{n,n+p}>_0 = {x^{n+p+r}\over {(-g)^{n+p+r}}} {{(p+r)(n+p)}
\over{n+p+r}} = <\chi_{r, p+r} \chi_{n+p,n}>_0\label{chi20'}
\b

One can write down this result in a label--independent way as follows
\a
<\chi_{r_1, s_1} \chi_{r_2,s_2}>_0 = {x^\Sigma \over {(-g)^\Sigma}}
{{M(r_1,s_1)M(r_2,s_2)}\over \Sigma}\label{chi20}
\b
where $\Sigma= r_1+r_2=s_1+s_2$ and $M(r,s)= max(r,s)$.
This formula also holds when the two labels of $\chi$ coincide.

The procedure just outlined for two--point functions works in general.
If we denote by $\chi^l$ a generic product of $l$ discrete states, one
first obtains $<\chi_{r,0} \chi^l>_0$ by suitably differentiating
(\ref{W1r-r0}), then one derives $<\chi_{r+p,r}\chi^l>_0$ by suitably
differentiating (\ref{W1rn0}). For the generic three point functions one
gets
\a
<\chi_{r_1,s_1}\chi_{r_2,s_2}\chi_{r_3,s_3}>_0 = {x^{\Sigma-1}\over
{(-g)^\Sigma}} M(r_1,s_1)M(r_2,s_2)M(r_3,s_3)\label{chi30}
\b
where $\Sigma = r_1+r_2+r_3=s_1+s_2+s_3$.

For the n--point functions with ${\rm n}>3$, there is more than one
possibility.
To explain this point we will consider for a while only n--point
functions $<\chi_{r_1,s_1} \ldots \chi_{r_n,s_n}>_0$ with $r_k \neq s_k$
for $k=1,\ldots ,n$ (the cases of coincident indices of $\chi$ are
obtained as limiting cases), and define their signature to be
$(p,q)$, where $p$ is the number of label $r_k>s_k$ and $q$ is the number
of labels $r_k<s_k$. For n--point functions with signature
$(1, n-1)$ we have the general formula
\a
&&<\chi_{r_1,s_1}\ldots \chi_{r_n,s_n}>_0 =
{x^{\Sigma -n+2}\over (-g)^\Sigma}
M(r_1,s_1)\dots M(r_n,s_n)(\Sigma -1)\ldots(\Sigma -n +3)\0\\
&& \Sigma = r_1+\ldots+r_n = s_1 +\ldots +s_n = j_1+\ldots +j_n\label{chin0}
\b
if $\Sigma >n-2$, and vanishes otherwise.
Here we have expressed $\Sigma$ in terms of the more standard $sl_2$ $j$
labels.

For other signatures, although there are in principle no obstacles,
explicit formulas are more elaborate to work out. For example, the
4--point function with signature $(2,2)$ is given by
\a
&&<\chi_{r_1,s_1}\ldots \chi_{r_4,s_4}>_0 =
{x^{\Sigma -2}\over {(-g)}^\Sigma}
M(r_1,s_1)\dots M(r_4,s_4)(\Sigma -\mu -1)\label{chi40}\\
&& \Sigma = r_1+\ldots+r_4 = s_1 +\ldots +s_4, \qquad
\mu= min(|r_1-s_1|,\ldots,|r_4-s_4|)\0
\b
if $\Sigma >2$, and vanishes otherwise.
All the above formulas extend to the boundary where some $r_k=s_k$,
and coincide for overlapping values of the labels.

We remark that the above formulas have been derived for states $\chi_{r,s}$
with $r$ and $s$ not simultaneously vanishing. To obtain CF's involving $p$
insertions of
$Q\equiv \chi_{0,0}$, one has simply to differentiate
$p$ times with respect to $x$ the corresponding CF without $Q$ insertions.
For CF's containing only $Q$ insertions, see section 5.

To conclude this section let us remark that CF's of pure
tachyonic states in genus 0, calculated with various methods,
are numerous in the literature, \cite{GK},
\cite{GD}, \cite{DFKut}, \cite{MP},\cite{D},\cite{KlPas},
\cite{DMP}. Our results coincide with these up to overall numerical
factors. Also formula (\ref{chin0}) has a counterpart in the literature,
\cite{D}.

{\it We can conclude that our results either coincide or extend previous
ones about discrete states. From this point of view, therefore, our conjecture
that $\chi_{r,s}$ are representatives of the discrete states of
$c=1$ string theory is confirmed.}

\section{The dispersionless extended Toda lattice hierarchy}

\setcounter{equation}{0}
\setcounter{subsection}{0}

The extended two--matrix model, as we showed in the section 2,
is governed by the extended Toda lattice hierarchy. In genus zero,
the characteristic integrable structure is the dispersionless
extended Toda lattice hierarchy.
In this section we will give
the explicit form of this integrable hierarchy and we will
show how to obtain, by means of the flow
equations and the coupling conditions, the same CF's we
obtained in the previous section by means of the $W$ constraints.
The basic ingredient is the continuum counterpart of the relation
(\ref{ddZ}), which provides a concrete expression for one--point
function, then after solving the coupling conditions, and making
use of the dispersionless Toda lattice hierarchy, we can easily
calculate all the correlators (besides the CF's among only $Q$),
and recover the results in the previous section.

We proceed as in \cite{BX1}. First we define the continuum quantities
in the following way (for more details, see Appendix)
\def\rg{ g^{\rm ren} }
\a
x= \frac{n}{N}, \qquad g^{\rm ren}_{r,s} = \frac{g_{r,s} }{N} \0
\b
They are the {\it renormalized} cosmological constant and coupling constants.
Further we define
\a
F_0(x) = \lim_{N\to\infty}  \frac{F_N}{N^2},\qquad
\zeta = \lim_{N\to\infty}  I_+ \0
\b
where $F_0$ is the genus zero free energy. The second equation is merely
symbolic and simply means that $\zeta$ is
the continuum counterpart of $I_+$.
If we define a matrix $\rho = \sum_n n E_{n,n}$, it is easy to see that we have
\a
\relax [ I_+, \rho] = I_+, \label{FPd}
\b
How this equation enters the Toda lattice hierarchy is explained, for
example, in \cite{BX4}.
The continuum counterpart gives the following basic Poisson bracket
\a
\{ \zeta , x \} = \zeta. \label{FP}
\b
This allows us to establish the following correspondence
\a
N~[~~,~~] \Longrightarrow \{~~,~~\} \label{ctp}
\b
and, similarly,
\a
\frac{1}{N} {\rm Tr} = \frac{1}{N} \sum_{0}^{N-1} \Longrightarrow \int_0^x dx
 \label{ctrint}
\b
together with the replacements
\a
&&Q(1) \rightarrow L, \qquad Q(2) \rightarrow \tl; \0\\
&&P(1) \rightarrow M, \qquad P(2) \rightarrow \tm \0
\b
where
\a
L=\zeta + \sum_{l=0}^\infty a_l \zeta^{-l}, \qquad
\tl=\frac{R}{\zeta} + \sum_{l=0}^\infty \frac{b_l}{R^l} \zeta^l. \label{ltl}
\b
Here $a_l$ and $b_l$ are the continuum fields that replace the lattice fields
$a_l(i)$ and $b_l(i)$ of eq.(\ref{jacobi}).

We stress that the above replacements holds only in genus 0.
The relation (\ref{ctp}) and (\ref{ctrint})
suggest that matrix size $N$ plays the role of Plank
constant in a Euclidean theory. In fact we need a Wick rotation
$N \to 1/(i\hbar)$, in order to be able to compare our results with those
obtained from a Minkowski space theory.

Next we define the operation \footnote{Notice that, with respect to \cite{BX1},
we have slightly changed our notation, precisely $\tl$ and $\hs (\tl)$ have
interchanged their
role.}
\a
\hs(\zeta)=\frac{R}{\zeta}, \qquad \hs(f)=f, \quad \forall {~\rm function}
{}~~~f \label{sigma}
\b
and eqs.(\ref{P1},\ref{P2}) become
\a
&& M= \sum_{r=1}^\infty rt_{1,r} L^{r-1} + x L^{-1} +
 \sum_{r=1}^\infty {{\d F_0}\over {\d t_{1,r}}}L^{-r-1}\label{cm}\\
&&\hs({\widetilde M}) = \sum_{r=1}^\infty rt_{2,r} {\tilde L}^{r-1} + x
{\tilde L}^{-1} +
\sum_{r=1}^\infty {{\d F_0}\over {\d t_{2,r}}}{\tilde
L}^{-r-1}\label{ctm}
\b

After this preparation, we can make the following statements:

\noindent{\it The dispersionless version of the coupling
conditions (\ref{coupling}) is}
\a
M + \sum_{r,s\geq1}rg_{r,s}L^{r-1}\tl^s=0, \qquad
\hs(\tm) + \sum_{r,s\geq1} sg_{r,s}L^r \tl^{s-1}=0, \label{cecoupling}
\b

\noindent{\it The dispersionless limit of the extended 2-dimensional Toda
lattice integrable hierarchy (\ref{CC12}, \ref{CC22}) is}
\ai
&&{{\d L}\over {\d \rg_{r,s}}} = \{ L, ~~ (L^r\tl^s)_-\}, \label{distoda1}\\
&&{{\d \tl}\over {\d \rg_{r,s}}} = \{ (L^r\tl^s)_+, ~~\tl\}. \label{distoda2}
\bj
Here the subscript + denotes the part containing non--negative
powers of  $\zeta$, while -- indicates the complementary part.
The commutativity of the flows directly follows from its discrete version.
Parallel hierarchies can be obtained by applying the $\sigma$ operator
to these equations.

Next, the continuum version of (\ref{ddZ}) provides
{\it the link between the free energy and Lax operators}, i.e.
\a
\frac{\d}{\d \rg_{r,s}}F = \int_0^x\, (L^r\tl^s)_{(0)}(y)dy \label{ddf}
\b
where subscript `${}_{(0)}$' means that we select the coefficient of zero--th
power term of $\zeta$. We
emphasize that this formula is valid on the full phase space. It opens
a way to calculate CF's by simply differentiating both sides
with respect to the appropriate coupling constants and use eqs.(\ref{distoda1},
\ref{distoda2}). Therefore this equation
together with the integrable hierarchy and the coupling conditions
completely determines the correlators in genus zero.

Let us see this formalism at work. As a simple example we compute the CF's in
the small phase space ${\cal S}_0$. In such a case, from eqs.(\ref{cecoupling},
\ref{cm},\ref{ctm}), we find
\a
L \rightarrow {\zeta},\qquad \tl \rightarrow
{1\over{-g}} {x\over \zeta} \0
\b
Therefore
\a
<\chi_{r,s}>_0 = \frac{\d F}{\d \rg_{r,s}} = \int_0^x\, (L^r\tl^s)_{(0)}(y)dy
\rightarrow\delta_{r,s} {x^{r+1}\over {r+1}} {1\over {(-g)^s}}\0
\b
Then, from
\a
{{\d L}\over {\d \rg_{r,s}}} = \{ L, ~~ (L^r\tl^s)_-\}
\rightarrow \left\{ \matrix {s x^{s-1} \zeta^{r-s+1}(-g)^{-s}, \quad & s>r\cr
0,\quad & s\leq r \cr}\right.
\label{distoda01}
\b
and
\a
{{\d \tl}\over {\d \rg_{r,s}}} = \{ (L^r\tl^s)_+, ~~\tl\}
\rightarrow  \left\{ \matrix{r x^s \zeta^{r-s-1} (-g)^{-s-1}, \quad & r\geq s
\cr 0, \quad & r<s \cr}\right.\label{distoda02}
\b
we obtain
\a
&&<\chi_{r,s}\chi_{k,l}>_0 =
{{\d^2 F}\over {\d \rg_{r,s} \d \rg_{k,l}}}={{M(r,s)M(k,l)}\over {(r+k)}}
{x^{r+k}\over {(-g)^{r+k}} }\delta_{r+k,s+l}\0\\
&&<\chi_{r,s}\chi_{k,l}\chi_{n,m}>_0 =
{{\d^3 F}\over {\d \rg_{r,s} \d \rg_{k,l}\d \rg_{n,m}}}=
M(r,s)M(k,l)M(n,m)
{x^{r+k+n-1}\over {(-g)^{r+k+n}} }\delta_{r+k+n,s+l+m}\0
\b
and so on. I.e. we recover the results of the previous section
\footnote{The suggestion to represent the discrete states by means of monomials
of
$L$ and $\tl$ is also present in ref.\cite{Taka}.}.

This is perhaps the right place to make a comment on the relation between
the method used in this section and the one used in the previous section.
In the previous section we used the $W$ constraints to calculate
CF's. Generally speaking the $W$ constraints
contain all the available information of a system, if we exclude the $N$--
dependent constant mentioned after eq.(\ref{parti3}). The $W$ constraints
were derived by using both flow equations and coupling conditions. In this
section we have been using both flow equations and coupling conditions. So it
is not surprising that we find the same results. The only difference is that
in this section we have been using a continuum formalism, while in
the previous one we used the discrete formalism. In other words,
in the present section we have first taken $N$ to infinity (with the necessary
rescalings and renormalizations) and carried out the calculations afterwards.
In the previous section we first carried out the calculations with $N$ finite:
taking $N$
to infinity in the final results (with the necessary rescalings) leads to the
same renormalized results as in this section -- replacing $N$ with the symbol
$x$ as we did in the last section understands the passage from discrete
quantities to (renormalized) continuum ones.

\section{Connection with the Penner model}

\setcounter{equation}{0}
\setcounter{subsection}{0}

The model we studied in the previous sections, characterized by the small phase
space ${\cal S}_0$, has a topological nature.
Remarkable evidence of this is provided by its connection
with the Penner model, \cite{penner}. Such connection is to be found
in the famous $N$--dependent integration constant (not determined
by the $W$ constraints) which was mentioned after eq.(\ref{parti3}).

{}From eq.(\ref{ZR}) and eq.(\ref{Todaeq}) we obtain
\a
{{\d^2 }\over{\d t_{1,1} \d t_{2,1}}} \ln R_n =
{{\d^2 }\over{\d t_{1,1} \d t_{2,1}}} (F_{n+1}-2F_n
+F_{n-1}) ,\label{RF''}
\b
Where $F_n = \ln Z_n$.
After integrating twice with respect to $t_{1,1}$ and $t_{2,1}$,
we have
\a
\ln R_n = F_{n+1}-2F_n+F_{n-1} \label{RF'}
\b
We remark that this equation can be directly derived from the expression
of the partition function (\ref{parti1}).
Possible integration constants have been dropped since they are irrelevant
for the present problem (they provide analytic contributions to the free
energy).

At this point it is more convenient to shift to a continuum formalism
(see Appendix) by introducing
the continuum quantities
\a
x \equiv {n\over N}, \qquad \epsilon \equiv {1\over N},
\qquad F(x,\epsilon) \equiv  \frac{F_n}{N^2}, \qquad
\label{contv}
\b
as we did in the previous section (from now on the quantities we mention
are understood to be the renormalized ones). However now we are
interested not only
in genus 0, but also in the higher genera. As is well--known
the contributions to the free energy corresponding to different genera rescale
in different ways. The appropriate genus expansion for the free energy
is as follows
\a
F(x,\epsilon) = \sum_{h=0}^\infty F_h(x) \epsilon^{2h } \label{topexp}
\b
where $F_h$ is the $h$ genus contribution. We also set
\a
F_{n\pm1} = e^{\pm\epsilon\d_x} F(x,\epsilon) \0
\b
Thus the continuum version of the equation (\ref{RF'}) is
\a
\ln R(x) = \frac{1}{\epsilon^2}\Bigl[\exp(\epsilon\d_x)
+\exp(-\epsilon\d_x)-2\Bigl]F(x,\epsilon). \label{crf}
\label{RF}
\b
Once we know $R(x)$, we can solve this equation to obtain the free energy.

Now at the particular critical point ($g=-1$) we are considering, $R(x) =x$,
which
corresponds to the genus 0 contribution, while higher genus contributions of
$R$ vanish (see below, section 6). Therefore we can easily obtain
the recursion relations among the free energy in different genera
\a
\d^2_x F_0(x) = \ln R(x)  \0
\b
and
\a
\sum_{l=1}^n \frac{1}{(2l)!} \d^{2l}_x F_{n-l}(x) =0,
\qquad \forall n\geq1.\label{recF}
\b
Using these recursion relations we can obtain the free energy for any genus.
For example,
\a
F_1(x) = - {1\over 12} \ln x, \qquad F_2(x) = -{1\over 240}x^{-2}, \0
\b
and so on. In general we have
\a
F_h(x) = \frac{B_{2h}} {2h(2h-2)}x^{2-2h}, \qquad \forall h\geq2. \label{fg}
\b
This assertion can be very easily proved by means of the properties of
the Bernoulli numbers $B_{2l}$, in particular
\a
\sum_{l=1}^{n-1} \frac{B_{2l}}{2l(2l-2)!(2n-2l)!} = \frac{1}
  {(2n)!}\0
\b
Eq.(\ref{fg}) is the free energy of the Penner model.
We see that $F_h(x)(h\geq2)$ indeed have scales according to the power $2(1-h)$
while $F_0$ and $F_1$ exhibit logarithmic
scaling violation, which is a typical feature of $c=1$ string theory
coupled to gravity. The coefficients of the
powers of $x$ are topological numbers.
We recall that the (virtual) Euler characteristics of moduli space of
Riemann surfaces with $n$ punctures
in genus $h$ was computed by Harer and Zagier \cite{HZ} and rederived
by Penner,\cite{penner}, with Feynman graph techniques
\a
\chi_{h}^{(n)}= \frac{(-1)^n(2h-3+n)!(2h-1)} {n!(2h)!}B_{2h}\0
\b
Therefore
\a
F_h= \chi_h^{(0)} x^{2-2h},
\qquad
<Q^n>_h = n! \chi_{h}^{(n)} x^{2-2h-n}\label{Qn}
\b
In particular we see that $Q$, the operator coupled to $x$, is
to be interpreted as the puncture operator in the present theory \cite{DV}.

It is a common belief that $c=1$ string theory is a topological field theory
with primaries, puncture equation, recursion relations and
a Landau--Ginzburg formulation \cite{Taka},\cite{EK},\cite{GM},\cite{Oz}.
On the basis of our results, which are far more complete than the ones
preexisting in the literature, we have reconstructed many sparse elements
of such conjecture but no compelling evidence. For this reason we do not
discuss them here, but collect them in Appendix B.

\section{Correlation functions. Higher genera.}

\setcounter{equation}{0}
\setcounter{subsection}{0}

To calculate correlators in higher genus we can use the $W$--constraints
as we did for genus 0 in section 3. However this method becomes more
and more cumbersome as we pass to higher and higher genera.
Just for the sake of comparison we will see an example below. As was
anticipated in section 4, however, a skillful use of the coupling
conditions and the flow equations may produce surprisingly powerful
shortcuts. This is what we are going to show next: compact formulas for
some correlators at all genera.

\subsection{Exact correlators at arbitrary genus}

In ${\cal S}_0$ the Jacobi matrices $Q(1)$ and $Q(2)$ take a very simple form
\a
Q(1)=I_+, \qquad
Q(2)=\sum_{l=1}^\infty {l\over{-g}}E_{l,l-1}\label{Q1Q2}
\b
which implies in particular -- compare with (\ref{jacobi1}) --
\a
R_l = {l\over {-g}}\label{R}
\b
Eqs.(\ref{Q1Q2}) can be derived by solving the coupling conditions.
In ${\cal S}_0$, the coupling conditions (\ref{coupling}) are simply
\a
P(1) + g Q(2) =0, \qquad \quad {\bar{\cal P}}(2) + gQ(1)=0. \label{cs0}
\b
Furthermore, in ${\cal S}_0$, eq.(\ref{P1}) shows us that
\a
P_+(1)=0, \qquad\qquad P_{n,n-1}(1) = n\0
\b
Combining these equations with the first equality in eqs.(\ref{cs0}),
we get the same $Q(2)$ in eqs.(\ref{Q1Q2}). Similarly
we can obtain $Q(1)$. In the rest of this section, we will show that
eqs.(\ref{Q1Q2}), together with the extended Toda lattice hierarchy, are
sufficient to exactly determine the correlators.

\subsubsection{The 2-point correlator of purely tachyonic states}

The calculation of the correlation functions consists of the following
three steps.

\noindent
{\it 1. Evaluating the correlator on lattice}:
Using eq.(\ref{ddZ}) and eqs.(\ref{CC12}, \ref{CC22}), we immediately get
\a
{\d^2\over{\d t_{1,m}\d t_{2,n}}}\ln Z_j(g) = {\tr}\Bigl([Q^m(1),
Q^n(2)]\Bigl) \0
\b
Evidently it is non--vanishing only if $m=n$, and a simple calculation
shows that
\a
&&~~{\d^2\over{\d t_{1,n}\d t_{2,n}}}\ln Z_j(g) = \tr\Bigl([Q^n(1),
    Q^n(2)]\Bigl) = \sum_{l=j-n+1}^j \prod_{i=l}^{l+n-1} R_i \0\\
&& = \frac{1}{(-g)^n} \sum_{l=j-n+1}^j l(l+1)\dots (l+n-1) \label{nnh}
\b

\noindent
{\it 2. Shifting to the continuum formalism}:
The {\it renormalized} correlation functions are
the derivatives of $F(x)$ with respect to the renormalized coupling
constants
\a
{\d^2\over{\d t_{1,n}^{\rm ren}\d t_{2,n}^{\rm ren}}} F(x,\epsilon)
 \equiv  < T_n T_{-n} >_{\rm ren} \0
\b
For simplicity, from now on, we drop the label $\scriptstyle{\rm ren}$ wherever
it should appear. From the
context it will be clear whether the quantities in question are
unrenormalized (before the continuum limit) or renormalized (after
the continuum limit).

A straightforward calculation shows
\a
< T_n T_{-n} >
&=& \frac{1}{(-g)^n} \sum_{l=1-n}^0 (x+l\epsilon)(x+(l+1)\epsilon) \ldots
 (x+(l+n-1)\epsilon) \label{tnepsilon}\\
 &=& \frac{1}{(-g)^n} \sum_{l=0}^{n-1} (x+l\epsilon)(x+(l-1)\epsilon) \ldots
(x+(l-n+1)\epsilon)\label{TTn}
\b

\noindent
{\it Separating the contributions from different genera}:
In order to single out the contributions from different genera, we recall
once again the topological expansion for the free energy
\a
F(x,\epsilon) = \sum_{h=0}^\infty F_h(x) \epsilon^{2h}
\b
Further  we observe that eq.(\ref{TTn})
correlator can be re--expressed as
\a
< T_n T_{-n} >
&=& \frac{1}{(-g)^n} \Bigl( \sum_{l=1-n}^0 e^{-l\epsilon\d_x} \Bigl)
 x(x+1\epsilon) \ldots (x+(n-1)\epsilon) \0\\
&=& \frac{1}{(-g)^n} \Bigl( \sum_{l=0}^{n-1} e^{l\epsilon\d_x} \Bigl)
 x(x-1\epsilon) \ldots (x-(n-1)\epsilon) \label{tt}\\
&=& \sum_{h\geq0} <T_n T_{-n}>_h \epsilon^{2h} \0
\b
The second equality exhibits the invariance under change
$\epsilon\to-\epsilon$, which amounts to the vanishing of the odd powers
of $\epsilon$ in the last expression. In other words,
the CF's have the same kind of expansion as the free energy.

Now let us define two series of numbers
\a
&& \al_i(n) = \sum_{l=0}^{n-1} l^i, \qquad i\geq0; \0\\
&& \beta_i(n) = \sum_{1\leq l_1<l_2<\ldots<l_i\leq n-1} \prod_{p=1}^i l_{p}
  \qquad i\geq0.\0
\b
The first few $\al_i(n)$ are
\a
&& \al_0(n) = n, \qquad \al_1(n) = \left(\bac n\\ 2\ea\right), \qquad
 \al_2(n) = \frac{2n-1}{3} \left(\bac n\\ 2\ea\right),\0\\
&& \al_3(n) = \frac{n^2(n-1)^2}{4}, \qquad
 \al_4(n) = \frac{(2n-1)(3n^2-3n-1)}{15} \left(\bac n\\ 2\ea\right)\0
\b
The first few $\beta_i(n)$ are
\a
&& \beta_0(n) = 1, \qquad \beta_1(n) = \left(\bac n\\ 2\ea\right), \qquad
\beta_2(n) = \frac{3n-1}{4} \left(\bac n\\ 3\ea\right),\0\\
&& \beta_3(n) = \left(\bac n\\ 2\ea\right) \left(\bac n\\ 4\ea\right), \qquad
\beta_4(n) = \frac{15n^3-30n^2+5n+2}{48} \left(\bac n\\ 5\ea\right)\0
\b
Furthermore, from the definition, we directly obtain
\a
\beta_{n-1}(n) = (n-1)!, \qquad \beta_i(n) = 0, \qquad \forall i\geq n.\0
\b
Notice that $b_i(n)\equiv\beta_i(n+1)$, where $b_i(n)$ were defined in
subsection 3.2. In terms of these numbers, we have the expansion
\a
x(x+\epsilon)(x+2\epsilon)\ldots (x+(n-1)\epsilon)
= \sum_{i=0}^n \beta_i(n) x^{n-i} \epsilon^i\0
\b
and
\a
\sum_{l=0}^{n-1} e^{-l\epsilon\d_x} = \sum_{l=0}^\infty \frac{(-1)^l}{l!}
\al_l(n) \epsilon^l \d_x^l\0
\b
Plugging these formulas into eq.(\ref{tt}), and remembering that terms with
the odd powers of $\epsilon$ vanish, we obtain the general two--point function
(which include the contributions from all the genera)
\a
< T_n T_{-n} > = \frac{1}{(-g)^n}\sum_{0\leq 2h\leq n} \epsilon^{2h}
\sum_{l=0}^{2h} (-1)^l \left(\bac n-l \\ 2h-l\ea\right)
\al_{2h-l}(n)\beta_l(n) x^{n-2h}\label{tt1}
\b
where $g$ is the renormalized coupling constant
 (we dropped the superscript for simplicity). The $h$--genus correlator is
\a
&& < T_n T_{-n}>_h = \frac{x^{n-2h}}{(-g)^n} \sum_{l=0}^{2h} (-1)^l
  \left(\bac n-l \\ 2h-l\ea\right)
  \al_{2h-l}(n)\beta_l(n), \qquad 2h< n; \label{tt2}\\
&& < T_n T_{-n}>_h =0, \qquad 2h\geq n. \0
\b
In particular we have
\ai
&& < T_n T_{-n}>_0 = \frac{1}{(-g)^n}nx^n,\label{ttg0} \\
&& < T_n T_{-n}>_1 = \frac{1}{(-g)^n}n \left(\bac n+1 \\ 4\ea\right) x^{n-2},
\label{ttg1} \\
&& < T_n T_{-n}>_2 = \frac{1}{(-g)^n}\frac{(3n^2-n-6)n}{8}
  \left(\bac n+1 \\ 6\ea\right) x^{n-4}.\label{ttg2}
\bj

\subsubsection{ The 3-point correlators of purely tachyonic states}

Now we turn our attention to the three--point correlation functions.
Once again, from the Toda lattice hierarchy (\ref{CC12}, \ref{CC22}) and
the relation (\ref{ddZ}), we get the correlator on the lattice
\a
{{\d^3 F_j}\over{\d t_{1,m}\d t_{1,n} \d t_{2,l}}} = \tr\Bigl(
[[Q^n(1),  Q^l_-(2)]_+, Q^m(1)]\Bigl)
+ \tr\Bigl( [Q^n_+(1), [Q^m(1),  Q^l_-(2)]]\Bigl)\label{<ttt>}
\b
Therefore in ${\cal S}_0$ we have to evaluate
\a
 \tr\Bigl( [I_+^n, [I_+^m, (RI_-)^l]]\Bigl)
&=& \delta_{l,n+m} \sum_{i=0}^{j-1}\Bigl(
  \prod_{p=i+1}^{i+l} R_p
 -\prod_{p=i-m+1}^{i+n} R_p - \prod_{p=i-n+1}^{i+m} R_p
+ \prod_{p=i-l+1}^{i} R_p \Bigl) \0\\
&=& \delta_{l,n+m} \Bigl( \sum_{i=j-m+1}^j -
  \sum_{i=j-l+1}^{j-n} \Bigl) \prod_{p=i}^{i+l-1} R_p \0
\b
After passing to the continuum, eq.(\ref{<ttt>}) becomes
\a
&& < T_nT_mT_{-l} > =
 \frac{\delta_{l,n+m}}{(-g)^l\epsilon}
  \Bigl( \sum_{p=0}^{m-1} e^{-p\epsilon\d_x}
- \sum_{p=n}^{l-1} e^{-p\epsilon\d_x} \Bigl)
x(x+\epsilon)(x+2\epsilon)\ldots (x+(l-1)\epsilon)\0\\
&&= - \frac{\delta_{l,n+m}}{(-g)^l\epsilon} \Bigl(
 \sum_{p=0}^{m-1} e^{p\epsilon\d_x}
- \sum_{p=n}^{l-1} e^{p\epsilon\d_x} \Bigl)
x(x-\epsilon)(x-2\epsilon)\ldots (x-(l-1)\epsilon) \0
\b
Again, the second equality confirms that terms with the odd powers of
$\epsilon$ vanish.
Expanding the two factors in terms of $\alpha, \beta$ numbers, we get the
compact formula for three point correlation functions
\a
&&<T_n T_m T_{-l}>_h =
\delta_{l,n+m} \frac{x^{l-2h-1}}{(-g)^l} \sum_{i=0}^{2h} (-1)^{i+1}
  \left(\bac l-i \\ l-2h-1\ea\right) \beta_i(n) \label{3ptach}\\\noal
&&\qquad \qquad\cdot
   \Bigl(\al_{2h-i+1}(n)+ \al_{2h-i+1}(m)-
   \al_{2h-i+1}(l)\Bigl), \quad 2h< l-1; \0\\\noal
&& < T_n T_m T_{-l}>_h =0, \qquad 2h\geq l-1. \0
\b
In particular for genus one we have
\a
<T_n T_m T_{-l}>_1 =\delta_{l,n+m}\frac{1}{(-g)^l} \frac{nm(n^2+m^2-l^2-2)}{4}
  \left(\bac l \\ 3\ea\right)x^{l-3}\label{3pg=1}\label{tttg1}
\b

Let us remark, after these two subsections, that {\it whenever two--
or three--point functions of tachyonic states in genus 1
are available in the literature in explicit form,
\cite{GK},\cite{KL},\cite{GD},
they coincide with the above results} up to overall normalization constants.

\subsubsection{Other discrete states}

The previous method does not work only for purely tachyonic states.
For example
\a
\frac{\d F_N}{\d g_{r,s}}=  \tr \Big(Q^r(1)Q^s(2)\Big)=
\delta_{r+s,0}\Big(Q^r(1)Q^r(2)\Big)\0
\b
On the other hand, in ${\cal S}_0$,
\a
\Big(Q^n(1)Q^n(2)\Big) = \sum_{p=0}^{N-1} R_{p+1}\ldots R_{p+n}
=\frac{1}{(-g)^n} \sum_{p=0}^{N-1} (p+1)\ldots (p+n)\label{chinnlat}
\b
The summation can be performed directly, and we obtain
\a
\frac{\d F_N}{\d g_{r,s}}
= n!\frac{1}{(-g)^n}\left(\bac n+N \\ n+1\ea \right)=
\frac{1}{(-g)^n} \frac {1}{n+1} N(N+1)\ldots (N+n)\0
\b
from which we recover the results of subsection 3.2. Now we can take
the continuum limit, which amounts to replacing $N$ by $x$, we finally have
\a
<\chi_{n,n}>_h = \beta_{2h}(n+1)
\frac{x^{n-2h+1}}{(n+1)(-g)^n}, \qquad 2h \leq n-1. \label{chinncon}
\b
In this particular case it is more convenient to carry out the calculation
in the discrete and take the continuum limit of the results, instead of
going to the continuum, as in the previous examples, and then carrying out the
calculations. The two methods lead to the same results as pointed out in
the Appendix.

Let us see another example: the two--point function for $\chi$ states.
\a
\frac{\d^2 F_N}{\d g_{r,s} \d g_{n,m}}
 &=& \tr \Big( [Q(1)^r , \Big(Q(1)^n Q(2)^m\Big)_-]
Q(2)^s \0\\
&&+~Q(1)^r [\Big(Q(1)^n Q(2)^m\Big)_+, Q(2)^s]\Big)\0
\b
Inserting eq.(\ref{Q1Q2}), we see immediately that this does not vanish only
if $ r+n = s+m$. Without loss of generality we can suppose $n\geq m$ and
$s\geq r$. Then we obtain
\a
\frac{\d^2 F_N}{\d g_{r,s} \d g_{n,m}}
  = \sum_{p=1}^N R_{p+s+m-1}\ldots R_p
-\sum_{p=1}^{N+r-s} R_{p+s-1}\ldots R_pR_{p+n-1}\ldots R_{p+n-m}\label{2chidis}
\b
To calculate this correlation function we proceed as in the example just shown.
We replace the explicit form of $R_p$, we notice that the genus $h$
contribution in proportional to $N^{m+s-2h}$ and single out the corresponding
coefficient. In the relevant formula we take the continuum limit, which amounts
to replacing $N$ with $x$, and obtain:
\a
&&<\chi_{r,s}\chi_{n,m}>_h = \frac {x^{m+s-2h}}{(-g)^{m+s}}\Bigg[ \frac
{b_{2h+1}(m+s)}{s+m+1} \label{2chi1}\\
&&~~~~~~~~~- \sum_{l=0}^{m+s}\gamma_l(s,n,m)\Bigg( \frac{1}{2}
\left(\bac s+m-l\\ 2h-l\ea\right)(r-s)^{2h-l}+
\frac{(r-s)^{1+2h-l}}{s+m-l+1}\left(\bac s+m-l+1\\ 1+2h-l\ea\right)\0\\
&&~~~~~~~~+\sum_{2\leq 2q \leq s+m-l} \frac{B_{2q}}{2q}
\left(\bac s+m-l\\ 2q-1\ea\right)
\left(\bac s+m-l-2q+1\\ 2h-l-2q+1\ea\right)(r-s)^{2h-l-2q+1}\Bigg)\Bigg]\0
\b
for $m+s > 2h$, and zero otherwise.
The convention $((-n)!)^{-1}=0$ is understood when $n>0$. In deriving
the above equation we have made use of a well--known summation formula
\cite{Gr}, which involves the Bernoulli numbers $B_{2q}$. Moreover we have
introduced
one more series of numbers
\a
\gamma_l(s,n,m) = \sum_{k=0}^l \sum_{0\leq i_1<i_2 \ldots <i_k\leq s-1}
\sum_{n-m\leq j_{k+1}<\ldots <j_l\leq n-1} i_1i_2\ldots i_k j_{k+1}\ldots j_l
\label{Cnumbers}
\b
which are non--vanishing only if $n>m$, and $l\leq m+s-1$. We have
in particular
\a
\gamma_0(n,m;s) &=& 1\0\\
\gamma_1(n,m;s) &=& {1\over 2}(s(s-1) + m(2n-m-1))\0\\
\gamma_2(n,m;s) &=& {1\over{24}} \Big(-2\,m - 3\,{m^2} + 2\,{m^3}
   + 3\,{m^4} + 12\,m\,n - 12\,{m^3}\,n - 12\,m\,{n^2} \0\\
   &+& 12\,{m^2}\,{n^2} - 2\,s + 6\,m\,s + 6\,{m^2}\,s
   -12\,m\,n\,s + 9\,{s^2}\0\\
    &-& 6\,m\,{s^2} - 6\,{m^2}\,{s^2} +
      12\,m\,n\,{s^2} - 10\,{s^3} + 3\,{s^4}\Big)\0
\b

We could of course continue with other more complicated examples. But
it is perhaps more interesting to make a comparison with the method
based on the $W$ constraints.

\subsection{Genus 1 correlators from $W$ constraints}.

In this subsection we resume the method of subsection 3.3, to calculate
correlation functions of discrete states at genus 1. This method is
more cumbersome and less efficient than the previous one. However it
provides an independent check of the previous calculations.
We proceed as in section 3.3 with an obvious
modification: in the expansion according to the
homogeneity degree (see Appendix), we have to take into account all the
contributions
up to and including the genus we are considering. For example, for genus 1,
in the expression for $T_n^{[r]}(1)$ we have to take into account the
following contributions
\a
&&T_n^{[r]}(1)= \sum_{\stackrel {k_1,k_2,\ldots,k_r}{l_1,l_2,\ldots,l_r}}
\Big(k_1k_2\ldots k_rg_{k_1,l_1}g_{k_2,l_2}\ldots g_{k_r,l_r}\frac {\d}
{\d g_{k_1+\ldots +k_r+n,l_1+\ldots +l_r}}\label{Tg1}\\
&&~~~~~~+{r\over 2}k_1k_2\ldots k_{r-1}(k_1+\ldots k_{r-1}-r+1)
g_{k_1,l_1}\ldots g_{k_{r-1},l_{r-1}}\frac{\d}{\d g_{k_1+\ldots+k_{r-1}+n,
l_1+\ldots+l_{r-1}}}\0\\
&&~~~~~~+\frac{r(r-1)}{6} k_1\ldots k_{r-2}\sum_{p=1}^{r-2}(k_p-1)(k_p-2)
g_{k_1,l_1}\ldots g_{k_{r-2},l_{r-2}}\frac{\d}{\d g_{k_1+\ldots+k_{r-2}+n,
l_1+\ldots+l_{r-2}}}\0\\
&&~~~~~~+\frac{r(r-1)}{4} k_1\ldots k_{r-2}
\sum_{\stackrel{p,q=1}{p<q}}^{r-2}(k_p-1)(k_q-1)
g_{k_1,l_1}\ldots g_{k_{r-2},l_{r-2}}\frac{\d}{\d g_{k_1+\ldots+k_{r-2}+n,
l_1+\ldots+l_{r-2}}}\Big)\0
\b
The first term in the RHS contains the genus 0 contribution.

For the sake of conciseness, we will not write down the corresponding
expression for ${\cal L}_n^{[r]}(1)$, which is even more complicated.
Eq.(\ref{Tg1}) gives an idea of the complications we run into by proceeding
with this method.

The first step of the calculation
consists in evaluating $<\chi_{n,n}>$, which we already know. Then by
differentiating $\Big({\cal L}_{-r}^{[r]}(1) - (-1)^r T_{-r}^{[r]}\Big) Z_N=0$
with respect to $g_{n+r,n}$ one obtains
\a
<\chi_{0,r}\chi_{n+r,n}>_1 = \frac{x^{n+r-2}}{24(-g)^{n+r}} r(n+r)(n+r-1)
(3n^2 +r^2 -n - r -2)\label{chi0rg1}
\b
Then differentiating
$\Big({\cal L}_{m}^{[r]}(1) - (-1)^r T_{m}^{[r]}(1)\Big) Z_N=0$ with $m>0$,
with
respect to $g_{n,n+m}$, and evaluating the result in ${\cal S}_0$,
we obtain
\a
&&<\chi_{n,n+m}\chi_{r+m,r}>_1 = \frac{x^{n+m+r-2}}{24 (-g)^{n+m+r}} \frac
{(n+m)(m+r)}{n+m+r-2}\cdot \label{chig1}\\
&&~~~~\cdot\Big(  -4 + 4\,m + 3\,{m^2} - 4\,{m^3} + {m^4} + 4\,n + 2\,m\,n -
6\,{m^2}\,n +
   2\,{m^3}\,n + 7\,{n^2} - 12\,m\,{n^2} \0\\
&&~~~~+ 4\,{m^2}\,{n^2} - 10\,{n^3} +
   6\,m\,{n^3} + 3\,{n^4} + 4\,r + 2\,m\,r - 6\,{m^2}\,r + 2\,{m^3}\,r \0\\
&&~~~~+ 10\,n\,r - 24\,m\,n\,r + 10\,{m^2}\,n\,r - 18\,{n^2}\,r +
   14\,m\,{n^2}\,r + 6\,{n^3}\,r + 7\,{r^2} - 12\,m\,{r^2} \0\\
&&~~~~ +  4\,{m^2}\,{r^2} - 18\,n\,{r^2} + 14\,m\,n\,{r^2} + 10\,{n^2}\,{r^2} -
   10\,{r^3} + 6\,m\,{r^3} + 6\,n\,{r^3} + 3\,{r^4}\Big)\0
\b
We have checked that this is exactly equal to eq.(\ref{2chi1}) above for $h=1$,
provided of course we choose corresponding labels.
As a particular case of (\ref{chig1}) or (\ref{chi0rg1}), we obtain
\a
<\chi_{m,0}\chi_{0,m}>_1= \frac{x^{m-2}}{24(-g)^m} (m+1)m^2(m-1)(m-2)\0
\b
which is eq.(\ref{ttg1}) above.

In the above calculation actually something new happens
with respect to section 3.3: there appear
`half--genus' contributions. In Appendix A we explain that $F$,
the free energy, is a sum of contributions from different genera, and $F_h$,
the genus $h$ contribution to the free energy scales like $2(1-h)$.
This is perfectly satisfactory for the ordinary two--matrix model.
But for the extended two--matrix we find also
contributions that scale like $(1-h)$, which we call with an effective
term `half--genus' contributions. This phenomenon is apparent for example
in eqs.(\ref{LZ},\ref{TZ}) and has already been remarked in section 6.
Such contributions appear as intermediate
results in the calculation on CF's via the $W$--constraints
and are necessary for the consistency of the theory.
{}From a practical point of view one has to them into account and
simply select the integer genus contributions in the final results.
As a matter of fact such contributions are not new as they appear
for example in the intermediate steps of the calculation of CF's
in models corresponding to n-th KdV hierarchies or other hierarchies.
But in the present case they appear as contributions {\it to} the CF's
themselves. They may in fact represent a new and interesting phenomenon:
they may imply that the extended two--matrix model takes contributions
not only from compact Riemann surfaces, but also from Riemann
surfaces with boundary.

\section{ Beyond the critical point}

\setcounter{equation}{0}
\setcounter{subsection}{0}

Till now we have only worked in the phase space ${\cal S}_0$.
What happens if we enlarge it? In this section we briefly discuss
this point. For the sake of simplicity, we focus our attention
on genus zero calculations.

\subsection{Perturbing around ${\cal S}_0$ }

Let us Taylor expand the free energy around ${\cal S}_0$. The genus $h$
part takes the form
\a
F_h(g_{r,s}) &=& F_h({\cal S}_0)
+ \sum_{r,s} g_{r,s} \frac{\d F_h({\cal S}_0)}{\d g_{r,s} }
+ \sum_{r_1,r_2;s_1,s_2} g_{r_1,s_1} g_{r_2,s_2} \frac{\d^2 F_h({\cal S}_0) }
  {\d g_{r_1,s_1} \d g_{r_2,s_2}}
   +\ldots \0\\
&=& + F_h({\cal S}_0) +
\sum_{r=1}^\infty g_{r,r} <\chi_{r,r}>_h \label{pertfs0}\\
&&~~~ \sum_{r_1,r_2;s_1,s_2} \delta_{r_1+r_2,s_1+s_2} g_{r_1,s_1} g_{r_2,s_2}
<\chi_{r_1,s_1}\chi_{r_2,s_2}>_h+\ldots\0
\b
In the second equality, the first term has been given in section 5, the other
terms are obtained by plugging in the correlation
functions of the small space ${\cal S}_0$. The `dots' contains the
contributions from multi--point CF's.
The free energy can be divided into two parts: coupling--independent
and  coupling--dependent.
These two terms are quite different: the  coupling--independent
term is singular at $x=0$, the other is regular;
even more crucial the  coupling--dependent term are completely
determined by $W_{1+\infty}$ constraints, while the  coupling--independent
term is not derivable from $W_{1+\infty}$ constraints. In fact, as
we have seen in section 5, it
stems from the Toda lattice equation and the coupling conditions.
Finally it is just the coupling--independent term
that give rise to the logarithmic violation of the scaling property
in genus zero. One should always bear in mind the existence and the
difference between these two contribution to the partition function.

\subsection{ Correlators in ${\cal S}$.}

The small phase space ${\cal S}$ is defined as follows
\a
{\cal S} \equiv \{ g_{r,s} = 0, \qquad {\rm except} \qquad
g_{1,1}=g, \qquad g_{0,r}=t_{1,r},\qquad g_{r,0}= t_{2,r} \} \0
\b
In other words, we switch off the extra couplings.

At first we notice that
in the continuum limit the genus 0 generators
${\cal L}^{[r]}_n$ take a very simple
form. Remember eqs.(\ref{J},\ref{J1},\ref{J2}) and remark that
\a
Z^{-1} J(z)Z=\sum_{r=1}^\infty r t_{1,r} z^{r-1} + {x \over z} +
\sum_{r=1}^\infty z^{-r-1} {{\d F_0}\over {\d t_{1,r}}} \0
\b
which has the same form as $M(L)$. From (\ref{W1rn}) and (\ref{tnr0}),
we get therefore
\a
 \sum_{\stackrel {i_1,\ldots,i_m\geq 1}{j_1,\ldots,j_m\geq 1}}
 \Bigl( \prod_{r=1}^m i_r g_{i_r,j_r}\Bigl)
 \frac{\d F_0}{\d g_{i_1+\ldots+i_m+n-m,j_1+\dots+j_m} }
=\frac{ (-1)^m }{(n+1)(m+1)} \oint M^{m+1} d L^{n+1}. \label{tnm0}
\b
Simply restricting to ${\cal S}$ we obtain
the one--point function
\a
<\chi_{n,m}>_0 =\frac{1}{(n+1)(m+1)(-g)^m}\oint M^{m+1}(L)dL^{n+1}\label{1pchi}
\b
We remark that this formula is very similar to the period representation
of one--point function in Landau--Ginzburg representation \cite{eguchi2}.

In order to obtain the multi--point functions of the extra states,
we differentiate eq.(\ref{tnm0}) with respect to the extra
couplings and evaluate the result in ${\cal S}$. For example,
the two--point function of the extra states is
\a
<\chi_{k,l}\chi_{n,m}>_0 = \frac{1}{(n+1)(m+1)(-g)^m}
\ddg {k,l} \oint M^{m+1}(L)dL^{n+1}
- \frac{km}{g} <\chi_{n+k-1, m+l-1}>_0 \label{2pchi}
\b
The first term in the RHS means that one should first take the
derivative with respect to the coupling $g_{k,l}$
(using the dispersionless Toda hierarchy), and then
do integration.
With the terminology of the Liouville approach \cite{kkli} \cite{mli}, the
first term may be interpreted as ``{\it bulk}'' contribution, while the second
term can be viewed as a ``{\it contact}'' term of two operators.
Similarly we have the three--point function
\a
<\chi_{r,s}\chi_{k,l}\chi_{n,m}>_0 &=& \frac{1}{(n+1)(m+1)(-g)^m}
 \frac{\d^2}{\d g_{r,s} \d g_{k,l}} \oint M^{m+1}(L)dL^{n+1} \0\\
&-&  \frac{km}{g} <\chi_{r,s}\chi_{n+k-1, m+l-1}>_0
   - \frac{rm}{g} <\chi_{k,l}\chi_{n+r-1, m+s-1}>_0 \label{3pchi}\\
&-&   \frac{rkm(m-1)}{g^2} <\chi_{n+r+k-2, m+s+l-2}>_0 \0
\b
Without much difficulty, we can derive formulas for the most general CF of
extra states on the small space
\a
< \chi_{n,m} \prod_{\mu=1}^k \chi_{i_\mu,j_\mu} >_0
&=& \frac{1}{(n+1)(m+1)(-g)^m} \Bigl(\prod_{\mu=1}^k \ddg {i_\mu,j_\mu}\Bigl)
  \oint M^{m+1}(L)dL^{n+1} \0\\
&-&  \sum_{r=1}^k \sum_{\rho\in S_k} \frac{i_{\rho(1)}i_{\rho(2)}\ldots
   i_{\rho(r)} m(m-1)\ldots (m-r+1) }{g^r}  \label{multipchi}\\
& &  < \chi_{n+i_{\rho(r+1)}+\ldots+i_{\rho(k)}-r,m+j_{\rho(r+1)}+\ldots
   j_{\rho(k)}-r} \prod_{s=1}^r \chi_{i_{\rho(s)},j_{\rho(s)}} >_0 \0
\b
where $\rho$ is an element of the symmetric group $S_k$.
The correlation functions involving pure tachyonic states are nothing but
particular cases of the above formulas.

\subsection{Correlators in ${\cal S}_1$. }

The above formulas for CF's are implicit. In order to get explicit results
we have to solve the coupling conditions. This is particularly simple
if the small phase space is
\a
{\cal S}_1 \equiv \{ g_{r,s} =0, \qquad
{\rm except} \qquad g_{1,1}=g,\qquad t_{1,1},\qquad t_{2,1} \} \0
\b
In this case, from the coupling conditions (\ref{cecoupling}),
we can write down two dispersionless Lax operators
\a
L= \zeta + \frac{ t_{2,1} }{-g}, \qquad \tl =
 \frac{1}{-g}(\frac{x}{\zeta}+t_{1,1}) \0
\b
together with
\a
R = \frac{x}{(-g)} \0
\b
Plugging these expressions into (\ref{ctm}),(\ref{1pchi}) and
(\ref{2pchi}), we obtain
\a
<\chi_{n,m}>_0 = \sum_{r=0}^{ {\rm min} (n,m) }
\left(\bac n \\ r\ea\right)
\left(\bac m \\ r \ea\right)\frac {x^{r+1} ~t_{1,1}^{m-r} ~t_{2,1}^{n-r}}
{(r+1)(-g)^{n+m-r} } \label{1pchis1}
\b
and
\a
&& <\chi_{k,l}\chi_{n,m}>_0 = km \sum_{r=1}^{ {\rm min} (n+k,m+l) }
    \left(\bac n+k-1 \\ r-1\ea\right)
    \left(\bac m+l-1 \\ r-1 \ea\right)\frac {x^r ~t_{1,1}^{m+l-r}
    ~t_{2,1}^{n+k-r}} {r(-g)^{n+m+k+l-r} } \label{2pchis1}\\
&&+ \sum_{\stackrel {0\leq i\leq n, 0\leq j\leq m}
    {0\leq r\leq k, 0\leq s\leq l} } (i-j)\theta(i-j)\delta_{i+r,j+s}
   \left(\bac n \\ i\ea\right)
    \left(\bac m \\ j \ea\right)
   \left(\bac k \\ r\ea\right) \left(\bac l \\ s\ea\right)
 \frac{  x^{j+s} ~t_{1,1}^{m+l-j-s} ~t_{2,1}^{n+k-i-r} }
   { (-g)^{n+m+k+l-i-r} }\0
\b
and so on. As for pure tachyons we find, for example,
\a
<T_n>_0= \frac{x t_{2,1}^n}{(-g)^n}, \qquad
<T_{-n}>_0= \frac {x t_{1,1}^n}{(-g)^n} \label{1tachyon}
\b
and
\a
<T_nT_m>_0=0, ,\quad <T_{-n}T_{-m}>_0=0, \quad
<T_n T_{-m}>_0 = \frac{nm}{-g} <\chi_{n-1,m-1}>_0 \label{2tachyon}
\b
as well as
 \a
&&<Q T_n T_m>_0 = 0, \qquad\qquad <QT_{-n}T_{-m}>_0 = 0 \0\\
&&<Q T_n T_{-m}>_0 = nm\sum_{r=1}^{ min(n,m) }
  \left(\bac n-1 \\ r-1 \ea\right)
  \left(\bac m-1 \\ r-1 \ea\right)
 \frac{ x^{r-1} t_{1,1}^{m-r} t_{2,1}^{n-r} }{(-g)^{n+m-r}} \label{metrics1}\\
&&<T_nT_mT_l>_0 = 0, \qquad
<T_nT_mT_{-l}>_0 = \frac{nml(l-1)}{(-g)^2}<\chi_{n+m-2,l-2}>_0 \0\\
&&<T_{-n}T_{-m}T_{-l}>_0 = 0, \qquad
  <T_nT_{-m}T_{-l}>_0 = \frac{n(n-1)ml}{(-g)^2}<\chi_{n-2,m+l-2}>_0. \0
\b
These correlators explicitly show that the selection rule implied by
Lemma 2, section 3, does not exist any more in ${\cal S}_1$.

{}From the above it is evident that the model ${\cal S}_1$ is quite different
from ${\cal S}_0$. However one can still conclude that
$Q$ plays the role of
puncture operator. Let us see this point in more detail.
\def\tg{{\widetilde g}}
We denote the couplings which are not in the small space ${\cal S}_1$
by $\tg_{r,s}$, and we suppose that they are infinitesimal
parameters. Then we can evaluate the free energy as follows
\a
F_h(\tg_{r,s}) &=& F_h({\cal S}_1)
+ \sum_{r,s} \tg_{r,s} \frac{\d F_h({\cal S}_1)}{\d \tg_{r,s} }
+ \sum_{r_1,r_2;s_1,s_2} \tg_{r_1,s_1} \tg_{r_2,s_2}
  \frac{\d^2 F_h({\cal S}_1) }
  {\d \tg_{r_1,s_1} \d \tg_{r_2,s_2}}
   +\ldots \0\\
&=& F_h({\cal S}_0) + \delta_{h,0} t_1s_1x +
\sum_{r,s} \tg_{r,s} <\chi_{r,s}>_h
+ \sum_{r_1,r_2;s_1,s_2} \tg_{r_1,s_1} \tg_{r_2,s_2}
   <\chi_{r_1,s_1}\chi_{r_2,s_2}>_h
 +\ldots \0
\b
The first two terms are calculated from Toda lattice equation recursively
by noting that $R=x$, which shows that $F_h({\cal S}_1)$ receives
an additional contribution only in genus zero. The remaining
terms are proportion to the infinitesimal parameters $\tg_{r,s}$. Now
taking derivatives with respect to $x$, we obtain
\a
<Q^n>_h = \frac{\d^n F_h(x)}{\d x^n} + \delta_{n,1}\delta_{h,0} t_1s_1
+ \tg-{\rm dependent~~~terms} \0
\b
Returning to the space ${\cal S}_1$ by setting $\tg_{r,s}=0$,
we see that the multi--point functions of the operator $Q$
(besides $<Q>_0$ ) still represent the Euler characteristics of
the punctured Riemann surface, i.e. $Q$ still plays
the role of puncture operator.

\section{Conclusion and outlook}

In this paper we have often utilized two different methods to derive the same
results.
We have seen that our results compare very well with the existing ones in the
literature. {\it We think we have given convincing evidence that
the extended two--matrix model not only provides a realization
of the discrete states of the $c=1$ string theory, but is also a very
powerful tool for solving it}.

As a matter of fact we have not mentioned so far the approach based
on the Liouville dressing of matter vertex operators, which makes essential use
of the Coulomb gas representation \cite{Kl-Po},\cite{Dotsenko},
\cite{ohta}.
Comparisons with this approach and related problems have been discussed in
the literature we have cited. We do not have anything new to add to them.

We have not addressed the theme of the ground ring of $c=1$ string theory
either,
\cite{witten}. We believe
a further extension of our model is necessary for us to be able to
describe such ring. We will return to this elsewhere.

\section*{}

\setcounter{equation}{0}
\setcounter{subsection}{0}

\subsection*{Appendix A. Discrete and continous formalism.}

In this Appendix we discuss the relation between the discrete and
continuous formalism, which are both used in this paper, and explain
why and when they give coincident results.

The equations we have to solve to calculate CF's are discrete as far
as the lattice variable $N$ is concerned. We recall that $N$ is
the common size of the matrices $M_1$ and
$M_2$. As we show in the paper one can
carry out many calculations sticking to the discrete formalism. This is in
fact the most natural thing to do. One then takes $N\to \infty$ in the results
themselves.
But shifting to a continuous variable $x$ is often more comfortable and more
effective.
The passage from $N$ to $x$ is regulated by precise rules,
which are dictated by the scaling properties af the quantities in question.
The latter differ genus by genus for the free energy and all the quantities
connected with it (CF's, fields, etc.).

In order to split the contributions from different genera
we rescale the lattice length. Simultaneously we have to rescale all
the quantities according to their degree of homogeneity $[~\cdot~]$
\a
\relax[g_{r,s}] = \al+\beta r+\gamma s, \qquad [N]=\al,
\qquad [F_h] = 2(1-h) \al\0
\b
where $F_h$ is the $h$--genus contribution
of the free energy. Analogous genus expansions hold
for all the fields $a_l$, $b_l$ and $R$ of the theory.
The numbers $\al, \beta$ and $\gamma$ above are arbitrary.
For the purposes of this paper it is enough to choose $\al =1$ and
$\beta=\gamma=0$. If we denote by $n$ (instead of $N$)
the matrix size and imbed the lattice variable $n$ into a finite lattice of
size $N$, then the above choice corresponds to rescaling the
lattice by $N$ so that the lattice spacing becomes $\epsilon= 1/N$.
We then set $x= n/N$. This choice has the virtue that the
contributions to the free energy corresponding to neighboring genera
differ by $\epsilon^2$. We stick to it throughout the paper.
After the correct rescalings have been performed, we can consider
$x$ as a continuous variable if $n$ and $N$ are large
enough and proceed according to the rules of the continuous differential
calculus.

Occasionally, however, we avoid such pedantic procedure as to introduce $n$
and $N$ as above, and simply say that $N$ (the matrix size) is replaced by
the renormalized variable $x$.

Two remarks are in order. First, the homogeneity argument is very important in
our approach, and it is suggested by the Feynman diagram approach of
\cite{BIZ}. This means nothing but the fact that we look only for solutions
with the above homogeneity properties; in particular this means that
we are interested in genus--expanded solutions.

Second, passing to the continuous variable $x$ is, in many cases, not
strictly necessary for the approach to work.
As a matter of fact, since in our two--matrix model there is an
arbitrarily large number of couplings, one
should eventually take $N$ to infinity: this is particularly evident from the
calculations of section 6. However one can take the attitude that $N$ is
arbitrarily large but finite, and one can let $N\to \infty$ (with the
necessary rescalings) once all
the calculations have been carried out.
The fact is that the
equivalence we establish between $f_{n+1}$ and $ \exp (\epsilon \d) f(x)$
is (up to a possible rescaling)
an exact relation even in the discrete case, at least
as long as $f_n$ is a polynomial in $n$.
As long as the quantities we have to do with
are polynomials, the two methods
(discrete and continuous) give the same results.

In conclusion, except when transcendental functions are involved, we can follow
two courses: {\it either} we go to the continuum (with the necessary rescalings
and renormalizations) and then do the calculations, {\it or} we carry out
the calculations with finite $N$ and take the continuum limit (with the
appropriate rescalings) on the results.

It also follows from the above considerations that,
although in this paper we have
privileged the $N\to \infty$ limit, the correlation functions of the two
matrix--model studied here can be solved with $N$ fixed and finite. A model
with finite $N$ can be of interest in itself.

\subsection*{Appendix B. Topological field theory properties.}

The so--called $c=1$ string theory is believed to be representable as
a topological field theory. In this Appendix we collect some (non conclusive)
evidence of this conjecture. I.e. we try to envisage
the model ${\cal S}_0$ studied in section 3 (at $g=1$)
as a topological field theory model.
 In order to be able to make such
assertion one should be able to identify the puncture operator, the primary
fields and the
descendants and show that they define an appropriate metric and
satisfy a puncture equation and the
appropriate recursion relations.

We have already seen that the puncture operator has to be $Q$.
As primary fields we take $T_n \equiv \chi_{n,0}$ and
$T_{-n}\equiv \chi_{0,n}$, where $n$ is any natural number
and $T_0 \equiv Q$, while all the other $\chi_{n,m}$ are descendants.
We recall that $T_n$ and $T_{-n}$ were identified in \cite{BX1}
with the purely tachyonic states ${\cal T}_n$ and ${\cal T}_{-n}$,
respectively, of $c=1$ string theory.

In this Appendix we have to pay some attention to indices. Since,
in order to conform with the current notation, we have introduced
negative labels, we will explicitly point it out whenever the labels $r,s,...$
denote integers and not simply natural numbers.

The metric (in the topological field theory sense) is given by
\a
\eta_{k,l} = <QT_kT_l>_0 \label{metric}
\b
where $k$ and $l$ are integers. The only nonzero elements are
(at $g=-1$)
\a
\eta_{n,-n}=\eta_{-n,n}=<QT_nT_{-n}>_0 \equiv {\d \over {\d x}}<\chi_{n,0}
\chi_{0,n}>_0 = n^2 x^{n-1}, \quad \eta_{0,0}= x^{-1}\0
\b
This metric is non--degenerate, the inverse is $\eta^{k,l}$ with
\a
\eta^{n,-n} = \eta^{-n,n} = n^{-2} x^{-n+1},\qquad \eta^{0,0} =x \0
\b
while all the other elements vanish. The associativity condition for
the structure constants $C_{i,j,k}$, $i,j,k$ integers,
\a
\sum_{k,l}C_{i,j,k}\eta^{k,l}C_{l,p,q} = \sum_{k,l}C_{i,p,k}\eta^{k,l}C_{l,j,q}
\0
\b
is easily seen to be satisfied once we notice that the only nonvanishing
three--point functions among primaries are
\a
&&C_{n,m,-n-m} =C_{-n,-m,n+m}
= <T_{-n}T_{-m} T_{n+m}>_0 = nm (n+m)
x^{n+m-1}\label{Cijk}\\
&&C_{n,-m, m-n} = \left\{ \bac nm(n-m)x^{n-1},\quad n>m\\
                               nm(m-n)x^{m-1},\quad n<m\ea\right.\0
\b
beside $C_{0,n,m} \equiv \eta_{n,-n} \delta_{n+m,0}$. The primary fields
form the commutative associative algebra ${\cal A}$
\a
T_i T_j = \sum_k C_{i,j}{}^k T_k,\qquad C_{i,j}{}^k \equiv
\sum_l C_{i,j,l}\eta^{l,k}\0
\b
where $i,j,k,l$ are integers and $T_0\equiv Q$. To prove it one has to use
\a
&&C_{n,m}{}^{n+m} = \frac {nm}{n+m},\qquad C_{0,n}{}^n =1,
\quad C_{n,-n}{}^0 = n^2 x^n, \qquad C_{0,0}{}^0 =1 \0\\
&& C_{n,-m}{}^{n-m}= \left\{ \bac \frac{nm}{n-m} x^m, \quad n>m\\
                                  \frac {nm}{m-n} x^n, \quad n<m\ea\right.\0
\b
where $n,m\neq 0$, $C_{i,j}{}^k =C_{j,i}{}^k$, and the other structure
constants
vanish.

Let us come now to the recursion relations and puncture equations.
The {\it recursion relations} in ${\cal S}_0$ are
\a
<\chi_{r,s}\chi_{r_1,s_1}\chi_{r_2,s_2}>_0 = M(r,s) \sum_{l,k}<\chi_{r-1, s-1}
T_l>_0 \eta^{l,k} <T_k \chi_{r_1,s_1}\chi_{r_2,s_2}>_0\label{recrel}
\b
where the labels $k$ and $l$ are understood to be integers.
The proof is very simple. Suppose for example that $r\geq s$.
Then
\a
{\rm LHS} = r M(r_1,s_1)M(r_2,s_2) x^{r+r_1+r_2-1}\0
\b
when $r+r_1+r_2 = s+s_1+s_2$ and vanishes otherwise. On the other hand
\a
{\rm RHS} = r <\chi_{r-1,s-1} T_{s-r}>_0 \eta^{s-r,r-s}<T_{r-s}
\chi_{r_1, s_1}\chi_{r_2,s_2}>_0 = r M(r_1,s_1)M(r_2,s_2) x^{r+r_1+r_2-1}\0
\b
when $r+r_1+r_2 = s+s_1+s_2$, and vanishes otherwise. The same can be proven
for $r\leq s$.

The {\it puncture equations} are designed to connect the CF's of
of the type
\a
<Q \chi_{r_1,s_1} \chi_{r_2 ,s_2} \ldots \chi_{r_n,s_n}>_0,\0
\b
where the $\chi$'s are extra states,
with CF's including neighboring descendants of them.
For dimensional reason the latter can only be
$<\chi_{r_1,s_1} \ldots\chi_{r_i-1 ,s_i-1} \ldots \chi_{r_n,s_n}>_0$.
A  heuristic relation which does this job for the CF's (\ref{chin0}) is
the following
\a
&&<Q \chi_{r_1,s_1} \chi_{r_2 ,s_2} \ldots \chi_{r_n,s_n}>_0 =
\label{puncture}\\
&&~~~~~~~~~~~~~~~~~~~~~~~
\sum_{i=1}^n\frac{M(r_i,s_i)}{M(r_i-1,s_i-1)}\frac{\Sigma-1}{n}
<\chi_{r_1,s_1} \ldots\chi_{r_i-1 ,s_i-1} \ldots \chi_{r_n,s_n}>_0\0
\b
where $\Sigma = r_1+\ldots+r_n= s_1 +\ldots +s_n$. In fact the LHS is
\a
<Q\chi_{r_1,s_1} \chi_{r_2 ,s_2} \ldots \chi_{r_n,s_n}>_0=
x^{\Sigma -n +1}M(r_1,s_1)\ldots
M(r_n,s_n) (\Sigma-1)\ldots (\Sigma -n +2)\0
\b
On the other hand the
generic term in the RHS of (\ref{puncture}) contains
\a
<\chi_{r_1,s_1}\ldots \chi_{r_k-1,s_k-1}\ldots \chi_{r_n ,s_n}>_0 &=&
M(r_1,s_1)\ldots M(r_k-1, s_k-1)\ldots M(r_n,s_n)\cdot \0\\
&&~~~~\cdot(\Sigma-2)\ldots(\Sigma -n+2)x^{\Sigma-n+1}\0
\b
Summing all the contributions in the RHS of (\ref{puncture})
we obtain the equality with the LHS.

The trouble with this formula is that it does not extend, as it is, to other
CF's and to higher genus. Modifications of this formula can be figured out
in many cases. For example, for one point functions,
\a
<(1- e^{-Q})\chi_{r,r}> = r<\chi_{r-1,r-1}>\label{punctureallg}
\b
is exact and is clearly the generalization of (\ref{puncture}) to every genus.
But, unfortunately, we do not have a general compact
formula for all the cases.

If we accept this point not as an imperfection of our proof
but as a peculiarity of the puncture operator $Q$,
we can then regard
our model as a topological field theory
in which $Q\equiv T_0$ is the puncture operator, $T_n$ and $T_{-n}$ ($n$
natural)
are the primary fields, while $\chi_{n+k,k}$ and $\chi_{k,n+k}$, with $k$
positive, are, respectively, the descendants. In particular $\chi_{k,k}$
are the descendants of $Q$.

Before we pass to the LG formulation, a comment about the position
$g=-1$ we made throughout this Appendix is in order.
The reason for such position is that only for this value of the coupling do
puncture and recursion equations hold, as is easy to check. Alternatively
the coupling $g$ can be absorbed into a redefinition of $\chi_{r,s}$:
precisely $\chi_{r,s} \rightarrow (\sqrt {-g})^{r+s} \chi_{r,s}$, which can of
course be achieved by rescaling the couplings $g_{r,s}$. In both cases
the coupling $g$ disappears from the expressions of the CF's.

The Landau--Ginzburg formalism allows us to represent in a condensed form
the essential properties of a topological field theory by means of a
suitable (usually, polynomial) potential $W$.
In the case of the topological field
theory (let us call it this way) in question
we must somehow expect an extreme situation.
The reason is that ${\cal S}_0$ looks very much like an extension of the $A_k$
type LG models to a model with an infinite number of fields in both positive
and negative direction. So it seems natural to look for a LG model in
which the ring of primary fields ${\cal R}$ is the ring of all monomials
in both negative and positive powers of a certain variable
$\zeta$ and $Q$ is represented by the
identity. Let us call the latter ring ${\cal P}$.
Since we expect ${\cal R}= {\cal P}
/{\d W}$ we see that $W$ can only be a constant at the critical point (in
particular $W$ will not be singular).

We have such a potential at hand. Let us define
\a
W \equiv \widetilde L L \label{W}
\b
At the critical point we have $W=x$. Next we set
\a
\phi_{r,s} \equiv \frac {\d W}{\d g_{r,s}}({\cal S}_0)= M(r,s)  x^s
\zeta^{r-s},
\qquad \phi_{0,0} \equiv \frac {\d W}{\d  x}({\cal S}_0) =1\0
\b
and in particular
\a
\phi_n \equiv \phi_{n,0} = n \zeta^n, \qquad
 \phi_0 \equiv \Phi_{0,0} =1,\qquad
 \phi_{-n} \equiv \phi_{0,n} = n x^n \zeta^{-n},\label{LGrep1}
\b
We define the following map
\a
\phi_n \leftrightarrow T_n ,\qquad \phi_0 \leftrightarrow Q,\qquad
\phi_{-n} \leftrightarrow T_{-n}\0
\b
and claim that it is an isomorphism between the algebra ${\cal R}$ and the
the field algebra ${\cal A}$.
It is elementary to prove the isomorphism by checking the few
non--trivial cases.

On the basis of the above argument we should not expect
the formula for the three--point function in terms of the potential $W$
to be exactly the same as in the ordinary LG models. In fact it is not, but it
is very close to it
\a
<\phi_{r_1,s_1}\phi_{r_2,s_2}\phi_{r_3,s_3}>_0=
\oint \frac {\d \zeta}{\zeta} \frac{\phi_{r_1,s_1}\phi_{r_2,s_2}\phi_{r_3,s_3}}
{W}\label{3pLG}
\b
This reproduces exactly formula (\ref{chi30}) with the obvious correspondences
$\phi_{r_i, s_i} \rightarrow \chi_{r_i,s_i}$, $i=1,2,3$.

\end{document}